\documentclass[preprintnumbers,unsortedaddress,superscriptaddress,notitlepage,square,sort&compress,comma,numbers]{revtex4-1}
\usepackage{lipsum}
\usepackage{CJK}
\usepackage{amsfonts}
\usepackage{amsmath}
\usepackage{amssymb}
\usepackage[english]{babel}
\usepackage{graphicx}
\usepackage{epsfig}
\usepackage{bm}
\usepackage{verbatim}
\usepackage[utf8]{inputenc}
\usepackage{booktabs}
\usepackage{multirow}
\usepackage{subfigure}
\usepackage{adjustbox}
\usepackage{slashed}
\usepackage{xcolor}
\usepackage{upgreek}
\usepackage[colorlinks=true,urlcolor=red,citecolor=red]{hyperref}
\usepackage[font=small]{caption}
\usepackage{float}
\usepackage{blindtext}
\usepackage{placeins}
\usepackage{bbm}
\usepackage{comment}
\usepackage[T1]{fontenc}
\usepackage{textcomp}

\usepackage[colorinlistoftodos]{todonotes}

\usepackage{soul}

\sethlcolor{yellow} 

\usepackage{verbatim}

\usepackage{tikz}
\usetikzlibrary{patterns}

\usetikzlibrary{arrows,shapes}
\usetikzlibrary{trees}
\usetikzlibrary{matrix,arrows} 
\usetikzlibrary{positioning}
\usetikzlibrary{calc,through}
\usetikzlibrary{decorations.pathreplacing}  
\usetikzlibrary{decorations.pathmorphing}
\usetikzlibrary{decorations.markings}
\tikzset{
    vector/.style={decorate, decoration={snake}, draw},
provector/.style={decorate, decoration={snake,amplitude=2.5pt}, draw},
antivector/.style={decorate, decoration={snake,amplitude=-2.5pt}, draw},
    fermion/.style={draw=black, postaction={decorate},
        decoration={markings,mark=at position .55 with {\arrow[draw=black]{>}}}},
    fermionbar/.style={draw=black, postaction={decorate},
        decoration={markings,mark=at position .55 with {\arrow[draw=black]{<}}}},
    fermionnoarrow/.style={draw=black},
    gluon/.style={decorate, draw=black,
        decoration={coil,amplitude=4pt, segment length=5pt}},
    scalar/.style={dashed,draw=black, postaction={decorate},
        decoration={markings,mark=at position .55 with {\arrow[draw=black]{>}}}},
    scalarbar/.style={dashed,draw=black, postaction={decorate},
        decoration={markings,mark=at position .55 with {\arrow[draw=black]{<}}}},
    scalarnoarrow/.style={dashed,draw=black},
    electron/.style={draw=black, postaction={decorate},
        decoration={markings,mark=at position .55 with {\arrow[draw=black]{>}}}},
bigvector/.style={decorate, decoration={snake,amplitude=4pt}, draw},
}\usetikzlibrary{decorations.markings}
\tikzstyle{block} = [draw, rectangle, 
    minimum height=3em, minimum width=6em]

\newcommand{\crn}{\nonumber \\}

\newcommand{\be}{\begin{equation}}
\newcommand{\ee}{\end{equation}}
\newcommand{\bea}{\begin{eqnarray}}
\newcommand{\eea}{\end{eqnarray}}
\newcommand{\ben}{\begin{enumerate}}
\newcommand{\een}{\end{enumerate}}
\newcommand{\bit}{\begin{itemize}}
\newcommand{\eit}{\end{itemize}}
\newcommand{\al}{\alpha}
\newcommand{\la}{\lambda}

\newcommand{\ga}{\gamma}
\newcommand{\va}{\varphi}
\newcommand{\om}{\omega}

\newcommand{\fr}{\frac}

\newcommand{\bc}{\begin{center}}
\newcommand{\ec}{\end{center}}
\newcommand{\Ga}{\Gamma}

\newcommand{\var}{\vartheta}

\newcommand{\ka}{\kappa}
\newcommand{\La}{\Lambda}
\newcommand{\si}{\sigma}

\newcommand{\mathsym}[1]{}

\definecolor{caribbeangreen}{rgb}{0.0, 0.8, 0.6}
\definecolor{parisgreen}{rgb}{0.31, 0.78, 0.47}
\definecolor{veronica}{rgb}{0.63, 0.36, 0.94}

\newcommand{\stai}{ Subatomic Physics Research Group,
		Science and Technology Advanced Institute,\\
		Van Lang University, Ho Chi Minh City 70000, Vietnam}
\newcommand{\steh}{  Faculty of Applied Technology, School of  Technology,  Van Lang University, Ho Chi Minh City 70000, Vietnam}

\newcommand{\usm}{Departamento de F\'{\i}sica, Universidad T\'{e}cnica Federico Santa Mar\'{\i}a Casilla 110-V, Valpara\'{\i}so, Chile}
\newcommand{\cctval}{Centro Cient\'{\i}fico-Tecnol\'ogico de Valpara\'{\i}so, Casilla 110-V, Valpara\'{\i}so, Chile}

\DeclareUnicodeCharacter{2212}{-}
\input{tcilatex}
\begin{document}

\title{Fermion masses and mixings and charged lepton flavor violation in a 3-3-1 model with inverse seesaw}

\author{A. E. C\'arcamo Hern\'andez}
\email{antonio.carcamo@usm.cl}
\affiliation{\usm}
\affiliation{\cctval}
\affiliation{Millennium Institute for Subatomic Physics at the High-Energy Frontier, SAPHIR, Calle Fern\'andez Concha No 700, Santiago, Chile}

\author{D. T. Huong
}
\email{dthuong@iop.vast.ac.vn}
\affiliation{Institute of Physics, Vietnam Academy of Science and Technology, 10 Dao Tan, Ba Dinh, Hanoi, Vietnam}

\author{H. N. Long
}
\email{hoangngoclong@vlu.edu.vn}
\affiliation{\stai}
\affiliation{\steh}

\author{Daniel Salinas-Arizmendi}
\email{daniel.salinas@usm.cl}
\affiliation{\usm}
\affiliation{\cctval}
\date{\today}

\begin{abstract}
We present a extension of the 3-3-1 gauge model supplemented by an $A_4$ flavor symmetry and cyclic discrete symmetries, including $Z_2$, $Z_2'$, $Z_3$, $Z_4$, $Z_7$ and $Z_{10}$. The model successfully reproduces the observed SM fermion mass hierarchies and mixing patterns in quark and lepton sectors. The smallness of the active neutrino masses is explained through an inverse seesaw mechanism, enabled by the introduction of right-handed and sterile Majorana neutrinos. In the quark sector, flavor-changing neutral currents (FCNCs) arise at tree level exclusively for up-type quarks via the new heavy neutral gauge boson exchange, leading to strong constraints from $D^0$–$\bar{D}^0$ mixing. The charged lepton sector exhibits sizeable flavor-violating effects, especially in the $\mu \rightarrow e\gamma$ decay, mediated by loops involving heavy neutrinos, new charged gauge bosons as well as charged scalars. We perform a detailed numerical fit of fermion masses and mixing parameters and identify viable regions of parameter space consistent with experimental data on CKM and PMNS mixing matrices. The model predicts branching ratios for charged lepton flavor violating decays and $\mu$–$e$ conversion rates within the sensitivity of future experiments. 

\end{abstract}

\maketitle

\section{Introduction} \label{intro}  Despite the remarkable consistency of the Standard Model (SM)
predictions with the experimental data, several issues remain unresolved. One such issue is the nonzero neutrino masses, confirmed by various neutrino oscillation experiments~\cite{McDonald:2016ixn,Kajita:2016cak,deSalas:2020pgw}. Additionally, the SM fails to account for the observed dark matter relic abundance and the matter-antimatter asymmetry of the Universe~\cite{Planck:2018vyg}. Furthermore,the SM cannot explain the existence of three fermion generations, the quantization of the electric charge, and the hierarchical structure of charged fermion masses. These shortcomings necessitate a more comprehensive theory, of which the SM may be considered a low energy effective theory.  The SM also fails in to explain the significant disparity between quark and lepton mixing patterns. While the Cabibbo-Kobayashi-Maskawa (CKM) matrix \cite{PhysRevLett.10.531, Kobayashi:1973fv} quantifies the misalignment between up- and down-quark Yukawa couplings, exhibiting a hierarchical structure with dominant diagonal elements, the Pontecorvo-Maki-Nakagawa-Sakata (PMNS) matrix \cite{Maki:1962mu} deviates significantly from diagonality.  The discovery of neutrino oscillations has established the existence of lepton flavor violation. However,the SM predicts negligible rates for such processes \cite{Petcov:1976ff,Marciano:1977wx,PhysRevLett.38.937,PhysRevD.16.1444}, like $\mu \to e \gamma $. To address this discrepancy and other shortcomings, various theories beyond the Standard Model (BSM) have been proposed. These BSM models, such as low-scale seesaw models, often predict enhanced rates for LFV decays, bringing them within the reach of current and future experiments.\\ 

The assumption of underlying BSM physics to explain the SM fermion mass hierarchies and mixing
patterns \cite
{Ma:2001dn,He:2006dk,Feruglio:2008ht,Feruglio:2009hu,Chen:2009um,Varzielas:2010mp,Altarelli:2012bn,Ahn:2012tv,Memenga:2013vc,Felipe:2013vwa,Varzielas:2012ai, Ishimori:2012fg,King:2013hj,Hernandez:2013dta,Babu:2002dz,Altarelli:2005yx,Gupta:2011ct,Morisi:2013eca, Altarelli:2005yp,Kadosh:2010rm,Kadosh:2013nra,delAguila:2010vg,Campos:2014lla,Vien:2014pta,Joshipura:2015dsa,Hernandez:2015tna,Karmakar:2016cvb,Borah:2017dmk,Chattopadhyay:2017zvs,CarcamoHernandez:2017kra,Ma:2017moj,CentellesChulia:2017koy,Bjorkeroth:2017tsz,Srivastava:2017sno,Borah:2017dmk,Belyaev:2018vkl,CarcamoHernandez:2018aon,Srivastava:2018ser,delaVega:2018cnx,Borah:2018nvu,Pramanick:2019qpg,CarcamoHernandez:2019pmy,CarcamoHernandez:2019kjy,Ding:2019zxk,Okada:2019uoy} may provide valuable clues to the nature of new physics (NP). The extension of the SM gauge symmetry to a higher weak isospin symmetry $SU(3)_L$, is a well-known framework for addressing the number of fermion generations, commonly referred to as the 3-3-1 model when including color and electric charges \cite{Pisano:1992bxx,PhysRevLett.69.2889,PhysRevD.22.738,Montero:1992jk}.  In addition to explaining the number of fermion families, this
new gauge principle offers potential solutions to fundamental questions such as electric charge
quantization \cite{Doff:1998we,deSousaPires:1998jc,deSousaPires:1999ca,Dong:2005ebq}, the strong CP conservation
\cite{Dong:2012bf,Pal:1994ba,Dias:2003zt}. Furthermore, it naturally addresses issues related to neutrino mass generation \cite{Tully:2000kk,Dias:2005yh,Chang:2006aa,Dong:2006mt,Dong:2008sw,Dong:2010gk,Dong:2010zu,Dong:2011vb,Boucenna:2014ela,Boucenna:2014dia,Boucenna:2015zwa,Okada:2015bxa,Pires:2014xsa,Dias:2010vt,Huong:2016kpa,Reig:2016tuk}, flavor physics \cite{GomezDumm:1994tz,Buras:2012dp,Buras:2013dea,Gauld:2013qja,Buras:2014yna,Buras:2015kwd,Dong:2015dxw,Binh:2024lez,Thu:2023xai,Duy:2022qhy,NguyenTuan:2020xls,Huong:2020csh}, dark matter stability \cite{Fregolente:2002nx,Hoang:2003vj,Filippi:2005mt,deS.Pires:2007gi,Mizukoshi:2010ky,Alvares:2012qv,Profumo:2013sca,Kelso:2013nwa,daSilva:2014qba,Dong:2013ioa,Dong:2014esa,Dong:2015rka,Dong:2013wca,Dong:2014wsa,Dong:2015yra,Huong:2016ybt,Alves:2016fqe,Ferreira:2015wja,Dong:2017zxo}, and even cosmic inflation and baryon asymmetry \cite{Huong:2015dwa,Dong:2018aak,Dong:2017ayu}. 
 
The $A_4$ discrete flavour group
has been used widely in the literature, due to its high predictive power and
since it provides a nice description of the SM fermion mass and mixing
pattern \cite
{Ma:2001dn,He:2006dk,Feruglio:2008ht,Feruglio:2009hu,Chen:2009um,Varzielas:2010mp,Altarelli:2012bn,Ahn:2012tv,Memenga:2013vc,Felipe:2013vwa,Varzielas:2012ai, Ishimori:2012fg,King:2013hj,Hernandez:2013dta,Babu:2002dz,Altarelli:2005yx,Gupta:2011ct,Morisi:2013eca, Altarelli:2005yp,Kadosh:2010rm,Kadosh:2013nra,delAguila:2010vg,Campos:2014lla,Vien:2014pta,Joshipura:2015dsa,Hernandez:2015tna,Karmakar:2016cvb,Borah:2017dmk,Chattopadhyay:2017zvs,CarcamoHernandez:2017kra,Ma:2017moj,CentellesChulia:2017koy,Bjorkeroth:2017tsz,Srivastava:2017sno,Borah:2017dmk,Belyaev:2018vkl,CarcamoHernandez:2018aon,Srivastava:2018ser,delaVega:2018cnx,Borah:2018nvu,Pramanick:2019qpg,CarcamoHernandez:2019pmy,CarcamoHernandez:2019kjy,Ding:2019zxk,Okada:2019uoy,Shimizu:2011xg,Kobayashi:2018scp}.  We adopt the $A_4$ group as discrete flavor symmetry of the $SU(3)_C \times SU(3)_L \times  U(1)_X$  3-3-1 model, leveraging its unique three-dimensional irreducible representation to naturally accommodate the three fermion families. The spontaneous breaking of symmetries generates the SM fermion mass and mixing patterns.  The  absence of exotic electric charges in fermion sector allows for low-scale linear or inverse seesaw mechanism, generating tiny neutrino masses and potentialy testable sterile neutrino at the $SU(3)_L \times U(1)_X$ breaking scale. Our model employs the inverse seesaw mechanism to generate light neutrino masses, thus making the model testable at the colliders. 

The remainder of the paper is organized as follows. In Section \ref{model}, we present the proposed model. Section \ref{quarkmixings} analyzes the implications of the model for quark masses and mixings. Lepton masses and mixings are discussed in Section \ref{leptons}. The scalar and gauge sectors of the model are detailed in Section \ref{scalarandgaugesector}. 
Section \ref{FCNCs} contains an analysis and discussion of the flavor-changing neutral currents.
Section \ref{LFV} derives the constraints on the model parameter space imposed by charged lepton flavor violating decays.
We conclude in section \ref{conclusions}.

\section{The model}
\label{model}
   We propose an extension of the economical 3-3-1 model with
three right handed Majorana neutrinos, where the $SU(3)_C\times
SU(3)_L\times U(1)_X$ gauge symmetry is supplemented by the $%
U(1)_{L_{g}} $ global lepton number symmetry and the $A_{4}\times
Z_{2}\times Z_{2}^{\prime }\times Z_{3}\times Z_{4}\times Z_{7}\times Z_{10}$ discrete group. In
this model, the scalar sector consists of two $SU(3)_L$ scalar
triplets and several gauge singlet scalars, introduced  to
generate viable and predictive SM fermion mass matrices that are consistent with the
low energy SM fermion flavor data. The scalar and fermionic spectra, along with
their transformations under the $SU(3)_C\times SU(3)_L\times
U(1)_X\times U(1)_{L_{g}}\times A_{4}\times Z_{2}\times Z_{2}^{\prime }\times Z_{3}\times
Z_{4}\times Z_{7}\times Z_{10}$ group, are presented in Tables \ref{tab:scalars}
and \ref{tab:fermions}, respectively. The dimensions of the $SU(3)_C$, $%
SU(3)_L$ and $A_{4}$ representations shown in Tables \ref{tab:scalars} and %
\ref{tab:fermions} are indicated by boldface numbers, while the
different $U(1)_X$ and $Z_N$ charges are written in additive notation.
Note that a field $\psi $ transforms under the $Z_N$ symmetry with a
corresponding $q_n$ charge as: $\psi \rightarrow e^{\fr{2\pi iq_n}{N}%
}\psi $, $n=0,1,2,3\cdots N-1$. We choose the $A_{4}$ symmetry because it is
the smallest nonabelian group possessing triplet and three singlet
irreducible representations (as detailed in Appendix~\ref{sec:appexA4}), thus allowing us to naturally accommodate the
three families of SM fermions. Additionally, the $A_{4}$, $Z_{3}$, $Z_{4}$
and $Z_{7}$ symmetries select the allowed entries of the SM down-type quark
mass matrices and determine the hierarchical pattern among these entries, which is necessary to correctly reproduce the observed SM fermion mass and mixing
structure. In addition to that, the $Z_{2}$ and $Z_{3}$ symmetries naturally explain the small values of the electron and down quark masses.
The $Z_{10}$ symmetry separates the $A_{4}$ scalar triplets
participating in the quark and neutrino Yukawa interactions from the ones
appearing in the charged lepton Yukawa terms, thus allowing one to treat these
sectors independently. Finally, the $Z_{2}^{\prime }$ symmetry is crucial for obtaining a
nearly diagonal sterile neutrino mass matrix with non-degenerate sterile
neutrinos.
In our model,
the masses for the SM charged fermions lighter than the top quark arise from non-renormalizable Yukawa interactions, after the
spontaneous breaking of the discrete symmetries. Consequently, the
spontaneous breaking of the $A_{4}\times Z_{2}\times Z_{2}^{\prime }\times
Z_{3}\times Z_{4}\times Z_{7}\times Z_{10}$ discrete group generates the observed 
pattern of SM fermion masses and mixing angles. The light active
neutrino masses are generated through an inverse seesaw mechanism, where the
smallness of its $\mu $ parameter is naturally explained by having higher
dimensional Yukawa interactions involving gauge singlet scalar fields
charged under the model's various discrete symmetries. Upon spontaneous breaking of these discrete symmetries, small lepton number violating Majorana mass terms are generated, leading to the $\mu$  parameter in the inverse seesaw mechanism. It is worth mentioning that the $%
U(1)_{L_{g}}$ global lepton number symmetry is spontaneously broken down to
a residual discrete $Z_{2}^{(L_{g})}$ by the vacuum expectation values
(VEVs) of the $U(1)_{L_{g}} $ charged gauge-singlet scalars $\xi _{i}$ ($%
i=1,2,3$), $\va  $ having a nontrivial $U(1)_{L_{g}}$ charge, as
indicated by Table \ref{tab:scalars}. The residual discrete $Z_{2}^{(L_{g})}$
lepton number symmetry, under which the leptons are charged and the other
particles are neutral, prevents the appearance of interactions involving an
odd number of leptons, which is crucial to guarantee the stability of the
proton. The corresponding massless Goldstone boson arising from the
spontaneous breaking of the $U(1)_{L_{g}}$ global symmetry, i.e., the
Majoron, is a $SU(3)_L$ scalar singlet, which does not cause any problems
in our model. Additionally, the fermions in our model do not exhibit exotic
electric charges. Consequently, the electric charge assumes the following form
\be
Q=T_{3}+\beta T_{8}+XI=T_{3}-\fr{1}{\sqrt{3}}T_{8}+XI,
\ee%
with $I=\text{diag}(1,1,1)$, $T_{3}=\fr{1}{2}\text{diag}(1,-1,0)$ and $T_{8}=\left(\fr{1}{2\sqrt{3}}\right)\text{diag}(1,1,-2)$ for a $SU(3)_L$ triplet. The $X$ parameter represents the charge associated with the $U(1)_X$ gauge group.
The lepton number is defined as \cite{Chang:2006aa,CarcamoHernandez:2017cwi,CarcamoHernandez:2019lhv,CARCAMOHERNANDEZ2024116588}:
\be
L=\fr{4}{\sqrt{3}}T_{8}+L_{g},
\ee
where $L_{g}$ is a conserved charge associated with the $U(1)_{L_{g}}$
global lepton number symmetry.
The full symmetry $\mathcal{G}$ features the following symmetry breaking pattern: 
\bea
&&\mathcal{G}=SU(3)_C\times SU\left( 3\right)_L\times U\left( 1\right)
_X\times U(1)_{L_{g}}\times A_{4}\times Z_{2}\times Z_{2}^{\prime }\times
Z_{3}\times Z_{4}\times Z_{7}\times Z_{10}{\xrightarrow{\La _{\text{int}}}}
\crn
&&\hspace{7mm}SU(3)_C\times SU\left( 3\right)_L\times U\left( 1\right)
_X\times U(1)_{L_{g}}{\xrightarrow{v_{\chi}}}  SU(3)_C\times SU\left( 2\right)_L\times U\left( 1\right)
_{Y}\times U(1)_{L_{g}} \crn
&&\hspace{7mm}{\xrightarrow{v_{\rho},v_{\varphi}, v_{\xi}}} SU(3)_C\times U\left( 1\right)
_{Q}\times Z_{2}^{(L_{g})}.
\label{break1}
\eea
To be consistent with the SM result, the symmetry breaking scales must satisfy the following hierarchy: $\Lambda_{\text{int}} >v_\chi \gg v_\rho,v_\varphi, v_\xi$. 
The $SU(3)_L$ triplet scalar fields $\chi $ and $\rho $ are represented
as:
\be \chi =%
\begin{pmatrix}
\frac{1}{\sqrt{2}} \left( \xi_\chi^\prime \pm i \zeta_\chi^\prime\right) \\
\chi _{2}^{-} \\
\fr{1}{\sqrt{2}}(v_{\chi }+\xi _{\chi }\pm i\zeta _{\chi })%
\end{pmatrix}%
,\hspace{1cm}\hspace{1cm}\rho =%
\begin{pmatrix}
\rho _{1}^{+} \\
\fr{1}{\sqrt{2}}(v_{\rho }+\xi _{\rho }\pm i\zeta _{\rho }) \\
\rho _{3}^{+}%
\end{pmatrix}%
\ee%
The $SU(3)_L$ fermionic triplets and antitriplets are given by:
\be
Q_{1L}=%
\begin{pmatrix}
u_{1} \\
d_{1} \\
J_{1} \\
\end{pmatrix}%
_L,\hspace{1cm}Q_{nL}=%
\begin{pmatrix}
d_n \\
-u_n \\
J_n \\
\end{pmatrix}%
_L,\hspace{1cm}L_{iL}=%
\begin{pmatrix}
\nu_{i_L} \\
l_{i_L} \\
\nu^{c}_{iR} \\
\end{pmatrix}%
,\hspace{1cm}n=2,3,\hspace{1cm}i=1,2,3.
\ee%
\begin{table}[th]
\begin{tabular}{|c|c|c|c|c|c|c|c|c|c|c|c|c|c|c|c|}
\toprule[0.2mm]
\hline
& $\chi $ & $\rho $ & $\si  _{1}$ & $\si  _{2}$ & $\si  _{3}$ & $\eta $
& $\zeta $ & $\var   $ & $\phi $ & $\Phi $ & $\Xi $ & $\xi $ & $\va  $
& $S_{1}$ & $S_{2}$ \\ \hline \hline
$SU(3)_C$ & $\mathbf{1}$ & $\mathbf{1}$ & $\mathbf{1}$ & $\mathbf{1}$ & $%
\mathbf{1}$ & $\mathbf{1}$ & $\mathbf{1}$ & $\mathbf{1}$ & $\mathbf{1}$ & $%
\mathbf{1}$ & $\mathbf{1}$ & $\mathbf{1}$ & $\mathbf{1}$ & $\mathbf{1}$ & $%
\mathbf{1}$ \\ \hline
$SU(3)_L$ & $\mathbf{3}$ & $\mathbf{3}$ & $\mathbf{1}$ & $\mathbf{1}$ & $%
\mathbf{1}$ & $\mathbf{1}$ & $\mathbf{1}$ & $\mathbf{1}$ & $\mathbf{1}$ & $%
\mathbf{1}$ & $\mathbf{1}$ & $\mathbf{1}$ & $\mathbf{1}$ & $\mathbf{1}$ & $%
\mathbf{1}$ \\ \hline
$U(1)_X$ & $-\fr{1}{3}$ & $\fr{2}{3}$ & $0$ & $0$ & $0$ & $0$ & $0$ & $%
0$ & $0$ & $0$ & $0$ & $0$ & $0$ & $0$ & $0$ \\ \hline
$U\left( 1\right) _{L_{g}}$ & $\fr{4}{3}$ & $-\fr{2}{3}$ & $0$ & $0$ & $%
0 $ & $0$ & $0$ & $0$ & $0$ & $0$ & $0$ & $2$ & $2$ & $0$ & $0$ \\ \hline
$A_{4}$ & $\mathbf{1}$ & $\mathbf{1}$ & $\mathbf{1^{\prime }}$ & $\mathbf{%
1^{\prime }}$ & $\mathbf{1^{\prime \prime }}$ & $\mathbf{3}$ & $\mathbf{3}$
& $\mathbf{3}$ & $\mathbf{3}$ & $\mathbf{3}$ & $\mathbf{3}$ & $\mathbf{3}$ &
$\mathbf{1^{\prime }}$ & $\mathbf{1^{\prime }}$ & $\mathbf{1^{\prime\prime}}$
\\ \hline
$Z_{2}$ & $1$ & $0$ & $0$ & $0$ & $1$ & $0$ & $0$ & $0$ & $0$ & $0$ & $0$ & $%
1$ & $1$ & $0$ & $0$ \\ \hline
$Z_{2}^{\prime }$ & $0$ & $0$ & $0$ & $0$ & $0$ & $0$ & $0$ & $0$ & $0$ & $0$
& $0$ & $0$ & $0$ & $1$ & $1$ \\ \hline
$Z_{3}$ & $0$ & $0$ & $1$ & $1$ & $0$ & $0$ & $0$ & $0$ & $0$ & $0$ & $0$ & $%
0$ & $0$ & $0$ & $0$ \\ \hline
$Z_{4}$ & $0$ & $0$ & $-1$ & $0$ & $0$ & $0$ & $0$ & $0$ & $0$ & $0$ & $0$ &
$2$ & $2$ & $0$ & $0$ \\ \hline
$Z_{7}$ & $0$ & $0$ & $0$ & $-1$ & $-3$ & $0$ & $0$ & $0$ & $0$ & $0$ & $0$
& $0$ & $0$ & $0$ & $0$ \\ \hline
$Z_{10}$ & $0$ & $0$ & $0$ & $0$ & $0$ & $0$ & $-5$ & $-5$ & $-3$ & $-1$ & $%
-1$ & $0$ & $0$ & $0$ & $0$ \\ \hline
\bottomrule[0.2mm]
\end{tabular}%
\caption{Scalar assignments under $SU(3)_C\times SU(3)_L\times
U(1)_X\times U\left( 1\right) _{L_{g}}\times A_{4}\times Z_{2}\times
Z_{2}^{\prime }\times Z_{3}\times Z_{4}\times Z_{7}\times Z_{10}$.}
\label{tab:scalars}
\end{table}
\begin{table}[th]
\begin{tabular}{|c|c|c|c|c|c|c|c|c|c|c|c|c|c|c|c|}
\toprule[0.2mm]
\hline
& $Q_{1L}$ & $Q_{2L}$ & $Q_{3L}$ & $u_{1R}$ & $u_{2R}$ & $u_{3R}$ & $D_{R}$
& $J_{1R}$ & $J_{2R}$ & $J_{3R}$ & $L_L$ & $l_{1R}$ & $l_{2R}$ & $l_{3R}$
& $N_{R}$ \\ \hline \hline
$SU(3)_C$ & $\mathbf{3}$ & $\mathbf{3}$ & $\mathbf{3}$ & $\mathbf{3}$ & $%
\mathbf{3}$ & $\mathbf{3}$ & $\mathbf{3}$ & $\mathbf{3}$ & $\mathbf{3}$ & $%
\mathbf{3}$ & $\mathbf{1}$ & $\mathbf{1}$ & $\mathbf{1}$ & $\mathbf{1}$ & $%
\mathbf{1}$ \\ \hline
$SU(3)_L$ & $\mathbf{3}$ & $\mathbf{\overline{3}}$ & $\mathbf{\overline{3}}
$ & $\mathbf{1}$ & $\mathbf{1}$ & $\mathbf{1}$ & $\mathbf{1}$ & $\mathbf{1}$
& $\mathbf{1}$ & $\mathbf{1}$ & $\mathbf{3}$ & $\mathbf{1}$ & $\mathbf{1}$ &
$\mathbf{1}$ & $\mathbf{1}$ \\ \hline
$U(1)_X$ & $\fr{1}{3}$ & $0$ & $0$ & $\fr{2}{3}$ & $\fr{2}{3}$ & $%
\fr{2}{3}$ & $-\fr{1}{3}$ & $\fr{2}{3}$ & $-\fr{1}{3}$ & $-\fr{1}{3%
}$ & $-\fr{1}{3}$ & $-1$ & $-1$ & $-1$ & $0$ \\ \hline
$U(1)_{L_{g}}$ & $-\fr{2}{3}$ & $\fr{2}{3}$ & $\fr{2}{3}$ & $0$ & $0$
& $0$ & $0$ & $-2$ & $2$ & $2$ & $\fr{1}{3}$ & $1$ & $1$ & $1$ & $-1$ \\
\hline
$A_{4}$ & $\mathbf{1^{\prime}}$ & $\mathbf{1^{\prime \prime}}$ & $\mathbf{1}$
& $\mathbf{1^{\prime}}$ & $\mathbf{1^{\prime}}$ & $\mathbf{1}$ & $\mathbf{3}$
& $\mathbf{1^{\prime}}$ & $\mathbf{1^{\prime \prime}}$ & $\mathbf{1}$ & $%
\mathbf{3}$ & $\mathbf{1^{\prime}}$ & $\mathbf{1^{\prime \prime}}$ & $%
\mathbf{1^{\prime}}$ & $\mathbf{3}$ \\ \hline
$Z_{2}$ & $1$ & $1$ & $1$ & $0$ & $0$ & $1$ & $0$ & $0$ & $0$ & $0$ & $0$ & $%
1$ & $0$ & $0$ & $0$ \\ \hline
$Z_{2}^{\prime }$ & $0$ & $0$ & $0$ & $0$ & $0$ & $0$ & $0$ & $0$ & $0$ & $0$
& $0$ & $0$ & $0$ & $0$ & $1$ \\ \hline
$Z_{3}$ & $1$ & $-1$ & $0$ & $1$ & $1$ & $0$ & $0$ & $1$ & $-1$ & $0$ & $0$
& $0$ & $-1$ & $1$ & $0$ \\ \hline
$Z_{4}$ & $0$ & $0$ & $2$ & $2$ & $0$ & $2$ & $2$ & $0$ & $0$ & $2$ & $1$ & $%
-1$ & $1$ & $-1$ & $1$ \\ \hline
$Z_{7}$ & $0$ & $0$ & $0$ & $4$ & $4$ & $0$ & $0$ & $0$ & $0$ & $0$ & $0$ & $%
0$ & $-3$ & $0$ & $0$ \\ \hline
$Z_{10}$ & $0$ & $0$ & $0$ & $0$ & $0$ & $0$ & $0$ & $0$ & $0$ & $0$ & $0$ &
$5$ & $3$ & $1$ & $0$ \\ 
\hline
\bottomrule[0.2mm]
\end{tabular}%
\caption{Fermion assignments under $SU(3)_C\times SU(3)_L\times
U(1)_X\times U\left( 1\right) _{L_{g}}\times A_{4}\times Z_{2}\times
Z_{2}^{\prime }\times Z_{3}\times Z_{4}\times Z_{7}\times Z_{10}$.}
\label{tab:fermions}
\end{table}
With the particle spectrum and symmetries specified in Tables \ref%
{tab:scalars} and \ref{tab:fermions}, we have the following relevant Yukawa
terms for both quark and lepton sectors:
\begin{equation} \label{Lyq}
\begin{aligned}
-\mathcal{L}_{Y}^{\left( q\right) } =& \ y_{1}^{\left( J\right) }\overline{Q}_{1L}\chi J_{1R}
+\sum_{n=2}^{3}y_n^{\left( J\right) }\overline{Q}_{nL}\chi^{\ast }J_{nR}
+y_{33}^{\left( u\right) }\overline{Q}_{3L}\rho ^{\ast}u_{3R}
+y_{23}^{\left( u\right) }\overline{Q}_{2L}\rho ^{\ast }u_{3R}\fr{\si  _{1}^{2}}{\La  ^{2}}\\
& +y_{13}^{\left( u\right)}\varepsilon _{abc}\overline{Q}_{1L}^{a}\left( \rho ^{\ast }\right)^{b}\left( \chi ^{\ast }\right) ^{c}u_{3R}\fr{\left( \si  _{1}^{\ast}\right) ^{2}}{\La  ^{3}}
+ y_{12}^{\left( u\right) }\varepsilon _{abc}\overline{Q}_{1L}^{a}\left( \rho ^{\ast }\right) ^{b}\left( \chi ^{\ast }\right)^{c}u_{2R}\fr{\left( \si  _{2}^{\ast }\right) ^{3}}{\La^{4}} +y_{22}^{\left( u\right) }\overline{Q}_{2L}\rho ^{\ast }u_{2R}\fr{\si
_{2}^{4}}{\La  ^{4}} \\
 & 
+
y_{11}^{\left( u\right)
}\varepsilon _{abc}\overline{Q}_{1L}^{a}\left( \rho ^{\ast }\right)
^{b}\left( \chi ^{\ast }\right) ^{c}u_{1R}\fr{\si  _{2}^{4}\si
_{1}^{2}}{\La ^{7}}  +\fr{1}{\La^2}y_{3}^{\left( d\right) }\varepsilon _{abc}\overline{Q}
_{3L}^{a}\rho ^{b}\chi ^{c}\left( \eta D_{R}\right) _{\mathbf{1}}
 \\
& +
y_{2}^{\left( d\right) }\varepsilon _{abc}\overline{Q}_{2L}^{a}\rho^{b}\chi ^{c}\left( \eta D_{R}\right) _{\mathbf{1}^{\prime \prime }}\fr{
\si  _{1}^{2}}{\La^{4}}
+
y_{1}^{\left( d\right) }\overline{Q}_{1L}\rho \left( \eta D_{R}\right) _{\mathbf{1}^{\prime }}\fr{\left( \si  _{1}^{\ast }\right) ^{2}\left( \si _{2}^{\ast }\right)^{3}\si  _{3}}{\La^{7}}
+H.c,
\end{aligned}
\end{equation}
\begin{equation}\label{Lyl}
\begin{aligned}
-\mathcal{L}_{Y}^{\left( l\right) } = & \ y_{1}^{\left( l\right) }\left(
\overline{L}_L\rho \zeta \right) _{\mathbf{1}}l_{1R}\fr{\si_{2}^{4}\si  _{1}^{2}\si  _{3}}{\La  ^{8}}
+y_{2}^{\left( l\right)}\left( \overline{L}_L\rho \phi \right) _{\mathbf{1}}l_{2R}\fr{\si_{2}^{4}}{\La  ^{5}}
+y_{3}^{\left( l\right) }\left( \overline{L}_L\rho\Phi \right) _{\mathbf{1}}l_{3R}\fr{\si  _{1}^{2}}{\La  ^{3}} 
\\
&+z_{1}^{\left( l\right) }\left( \overline{L}_L\rho \var   \right) _{%
\mathbf{1}}l_{1R}\fr{\si  _{2}^{4}\si  _{1}^{2}\si  _{3}}{\La
^{8}}+z_{3}^{\left( l\right) }\overline{L}_L\left( \rho \Xi \right) _{%
\mathbf{1}}l_{3R}\fr{\si  _{1}^{2}}{\La  ^{3}} \\
&+y_{1\chi }^{\left( L\right) }\left( \overline{L}_L\chi N_{R}\right) _{%
\mathbf{1^{\prime \prime }}}\fr{S_{1}}{\La  }+y_{2\chi }^{\left(
L\right) }\left( \overline{L}_L\chi N_{R}\right) _{\mathbf{1}^{\prime }}%
\fr{S_{2}}{\La  }+x_{\rho }\varepsilon _{abc}\left( \overline{L}%
_L^{a}\left( L_L^{C}\right) ^{b}\right) _{\mathbf{3a}}\left( \rho ^{\ast
}\right) ^{c}\fr{\eta \si  _{1}^{2}}{\La  ^{3}}  \\
&+y_{1N}\left( N_{R}\overline{N_{R}^{C}}\right) _{\mathbf{3s}}\xi \fr{%
\si  _{2}^{4}\si  _{1}^{2}\si  _{3}}{\La  ^{7}}+y_{2N}\left( N_{R}%
\overline{N_{R}^{C}}\right) _{\mathbf{1}}\va  \fr{\si  _{2}^{4}\si
_{1}^{2}\si  _{3}}{\La  ^{7}}+H.c.,
\end{aligned}
\end{equation}
where the subscripts $\mathbf{1},\ \mathbf{1^\prime},\ \mathbf{1^{\prime \prime}},\ \mathbf{3s},\ \mathbf{3a}$ denote the irreducible representations under the discrete symmetry group $A_4$. The corresponding tensor products are detailed in Appendix~\ref{sec:appexA4}.\\

\begin{figure}
\centering
\subfigure[]{\includegraphics[width=0.2\linewidth]{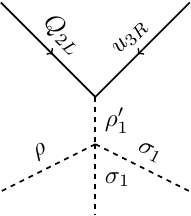}} \ \
\subfigure[]{\includegraphics[width=0.2\linewidth]{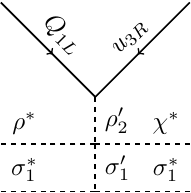}}\ \
\subfigure[]{\includegraphics[width=0.2\linewidth]{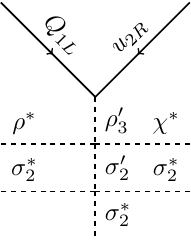}}\ \
\subfigure[]{\includegraphics[width=0.2\linewidth]{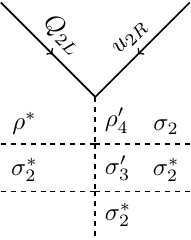}}\ \
\subfigure[]{\includegraphics[width=0.2\linewidth]{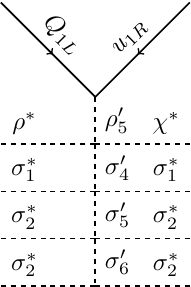}}\ \
\subfigure[]{\includegraphics[width=0.2\linewidth]{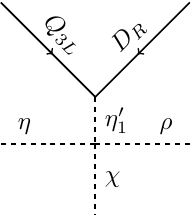}}\ \
\subfigure[]{\includegraphics[width=0.2\linewidth]{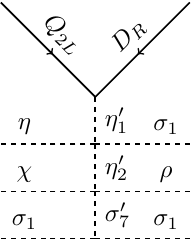}}\ \
\subfigure[]{\includegraphics[width=0.2\linewidth]{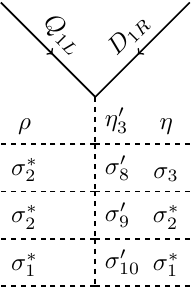}}
\caption{Feynman diagrams that induce the non-renormalizable
operators in quark sector.}
\label{fig:quarkdiagram}
\end{figure}

\begin{figure}
\centering
\subfigure[]{\includegraphics[width=0.2\linewidth]{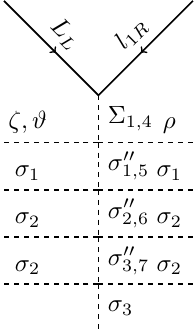}} \ \
\subfigure[]{\includegraphics[width=0.2\linewidth]{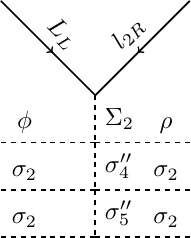}}\ \
\subfigure[]{\includegraphics[width=0.2\linewidth]{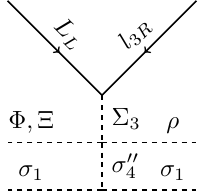}}\ \
\subfigure[]{\includegraphics[width=0.2\linewidth]{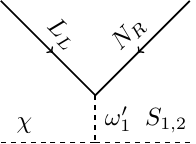}}\ \
\subfigure[]{\includegraphics[width=0.2\linewidth]{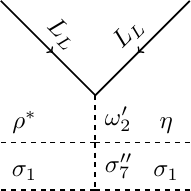}}\ \
\subfigure[]{\includegraphics[width=0.2\linewidth]{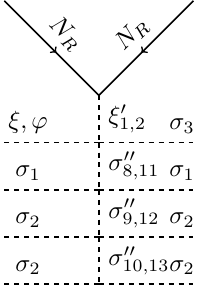}}
\caption{Feynman diagrams that induce the non-renormalizable
operators in lepton sector.}
\label{fig:leptondiagram}
\end{figure}

In our model, the Yukawa terms contain several non-renormalizable interactions, as observed in Eqs. \eqref{Lyq} and \eqref{Lyl}. These operators arise at high energies, where heavy mediators induce effective terms at low energies. They allow for an explanation of the fermion mass hierarchy without fine-tuning. Additionally, these operators may play a key role in generating light neutrino masses within the inverse seesaw mechanism. These operators have a general structure of the form%
\begin{equation}
\overline{\psi_L} H_1 H_2\Psi_R \frac{H_3^n}{\Lambda^{n+1}},
\end{equation}
where $\psi_L$ and $\Psi_L$ are fermions, $H_i$ are scalar fields that acquire VEVs, generating the observed hierarchy in the fermion masses.
The non-renormalizable operators could arise from a more fundamental theory with new scalar fields and heavy fermions. In particular, the effective operators in \eqref{Lyq} and \eqref{Lyl} may originate from the following renormalizable interactions
\begin{equation}
\overline{\psi_L} H_4 \tilde{\Psi}_R, 
\end{equation}
where the fermions $\tilde{\Psi}$ is heavy intermediate states that, once integrated out of the low-energy spectrum, induce the effective operators appearing in our model.
The Yukawa operators of non-normalizable up and down quarks are
has a possible ultraviolet origin of the non-renormalizable terms is that they are generated at low energies through the Feynman diagrams presented in Figure~\ref{fig:quarkdiagram}, where $H'_i$ are mediator fields with masses on the order of the model cutoff scale $\Lambda_{\text{int}}$. Meanwhile, the non-renormalizable lepton Yukawa operators can be generated at low energies from the Feynman diagrams shown in Figure~\ref{fig:leptondiagram}.

In order to comply with the current data on SM fermion masses and mixing
angles with less parameters than observables (especially in the lepton
sector), we consider the following VEV patterns for the $A_{4}$ triplets
gauge singlet scalars:
\bea
\left\langle \eta \right\rangle &=&\fr{v_{\eta }}{\sqrt{3}}\left(
1,1,1\right) ,\hspace{1cm}\hspace{1cm}\left\langle \xi \right\rangle =\fr{%
v_{\xi }}{\sqrt{3}}\left( 1,-1,1\right) , \\
\left\langle \zeta \right\rangle &=&\fr{v_{\zeta }\cos \al  }{\sqrt{3}}%
\left( \cos \al  -\sin \al  ,\cos \al  -\om \sin \al  ,\cos
\al  -\om ^{2}\sin \al  \right) ,\hspace{1cm}\left\langle \var
\right\rangle =\fr{v_{\var   }\om ^{2}\sin \al  }{\sqrt{3}}\left(
1,\om ^{2},\om \right) , \\
\left\langle \phi \right\rangle &=&\fr{v_{\phi }}{\sqrt{3}}\left( \cos
\al  +\sin \al  ,\om \cos \al  +\sin \al  ,\om ^{2}\cos \al
+\sin \al  \right) , \\
\left\langle \Phi \right\rangle &=&-\fr{v_{\Phi }\om \sin \al  }{%
\sqrt{3}}\left( \cos \al  -\sin \al  ,\cos \al  -\om \sin \al
,\cos \al  -\om ^{2}\sin \al  \right) \\
\left\langle \Xi \right\rangle &=&\fr{v_{\Xi }\cos \al  }{\sqrt{3}}%
\left( 1,\om ^{2},\om \right) ,\hspace{1cm}\om =e^{\fr{2\pi i}{3}%
},
\eea%
which are consistent with the scalar potential minimization equations for a
large region of parameter space.

Given that the breaking of the $A_{4}\times Z_{2}\times Z_{3}\times
Z_{4}\times Z_{7}\times Z_{10}$ discrete group produces the SM charged
fermion mass and quark mixing pattern, we set the VEVs of the $SU(3)_L$
singlet scalar fields with respect to the Wolfenstein parameter $\la
=0.225$ and the model cutoff $\La  $. If the gauge symmetry break as patterns given in \eqref{break1}, we have 
\be
v_{\rho}\sim v_{\xi}\sim v_{\va }\sim \la^{5}\La  \ll v_{\chi}\sim \la ^{2}\La < v_{\si_{i}}\sim v_{S_k}\sim v_{\eta}\sim v_{\zeta}\sim v_{\vartheta} \sim v_{\phi}\sim v_{\Phi }\sim v_{\Xi}\sim\la\La,\hspace{0.2cm}i=1,2,3,\hspace{0.2cm}k=1,2,
\label{VEVsinglets1}
\ee
where the model cutoff $\La  $ is the scale of the UV completion of the
model, which can corresponds to the mass scale of the Froggatt-Nielsen
messenger fields.

\section{Quark masses and mixings}
\label{quarkmixings}
The quark Yukawa terms of Eq. (\ref{Lyq}) give rise to the following SM mass
matrices for quarks:
\bea
M_{U} &=&\fr{v_\rho}{\sqrt{2}}\left(
\begin{array}{ccc}
c_{1}\la  ^{8} & b_{1}\la  ^{5} & a_{1}\la  ^{4} \\
0 & b_{2}\la  ^{4} & a_{2}\la  ^{2} \\
0 & 0 & a_{3}%
\end{array}%
\right) ,\hspace{1cm}\hspace{1cm}M_{D}=\fr{v_\rho}{\sqrt{2}}\left(
\begin{array}{ccc}
g_{1}\la  ^{7} & 0 & 0 \\
0 & g_{2}\la  ^{5} & 0 \\
0 & 0 & g_{3}\la  ^{3}%
\end{array}%
\right) R_{D},\hspace{1cm}\hspace{1cm}  \crn
R_{D} &=&\fr{1}{\sqrt{3}}\left(
\begin{array}{ccc}
1 & 1 & 1 \\
1 & \om ^{2} &  \om\\
1 & \om & \om ^{2}%
\end{array}%
\right),
\label{Mq}
\eea
where $c_{1}$, $b_n$ ($n=1,2$),$\ a_{i}$,$\ g_{i}$ ($i=1,2,3$) are $%
\mathcal{O}(1)$ dimensionless parameters, assumed to be real, excepting $%
a_{1}$, which is taken to be complex. Notice that we take $a_{1}$ to be
complex since the CP violating phase in the quark sector is associated with
the quark mixing angle in the $1$-$3$ plane, as indicated by the Standard
parametrization of the CKM
quark mixing matrix. In addition, notice that parameter $v_{\rho}=246$ GeV corresponds to the scale of electroweak symmetry
breaking. The SM quark mass matrices of Eq.~\eqref{Mq}, imply that the quark
mixing angles are generated from the up type quark sector.

Furthermore, we find that the exotic quark masses are given by:
\be
m_{J_{i}}=y_{i}^{\left( J\right) }\fr{v_{\chi }}{\sqrt{2}},\hspace{1cm}%
\hspace{1cm}i=1,2,3.  \label{mexotics}
\ee%
The experimental values of the quark masses \cite{Bora:2012tx,Xing:2007fb},
mixing angles and Jarlskog invariant \cite{Tanabashi:2018oca} are well
reproduced, as indicated in Table \ref{quarkobs}. Consequently, the fit for the quark sector
was performed by minimizing a $\chi^2_q$ function, defined as, 
\begin{equation} \label{eq:chiquarkfit}
\chi^2_q=\sum_i\left(\frac{O_i^{\text {model}}-O_i^{\exp }}{\sigma_i^{\exp }}\right)^2  
\end{equation}
where $ O_i = \{ m_i, \sin \theta_{i j}^{(q)}, J_q\}$, where $m_i$ are the masses of the quarks $(i=u, c, t, d, s, b)$, $\sin \theta_{i j}^{(q)}$ is the sine function of the quark mixing angles (with $j, k=1,2,3$ ) and $J_q$ is the quark Jarlskog invariant. The supra indices represent the experimental (exp) and theoretical (model) values, and the $\sigma$ are the experimental errors. Obtaining the following values for our best-fit point
for the following values of the quark sector parameters:
\begin{equation}\label{Benchmark1}
\begin{array}{l l l l l l l l}
 c_1  \simeq  -2.09 \pm 0.13, & b_1  \simeq 2.84 \pm 0.14, &  b_2  \simeq   2.77 \pm 0.13, & \left|a_1\right|  \simeq   3.63 \pm 0.18, & a_{2} \simeq   0.80 \pm 0.03,  \\
 a_{3} \simeq  0.99 \pm 0.03 , & g_{1}\simeq -0.89 \pm 0.04, &  g_{2}\simeq 0.95 \pm 0.06, &  g_{3}\simeq   -2.12 \pm 0.07, &  \arg \left(
a_{1}\right)  \simeq 23.35 \pm 1.57^{\circ }.
\end{array}
\end{equation}

\begin{figure}[H]
\centering
\subfigure[]{\includegraphics[scale=.45]{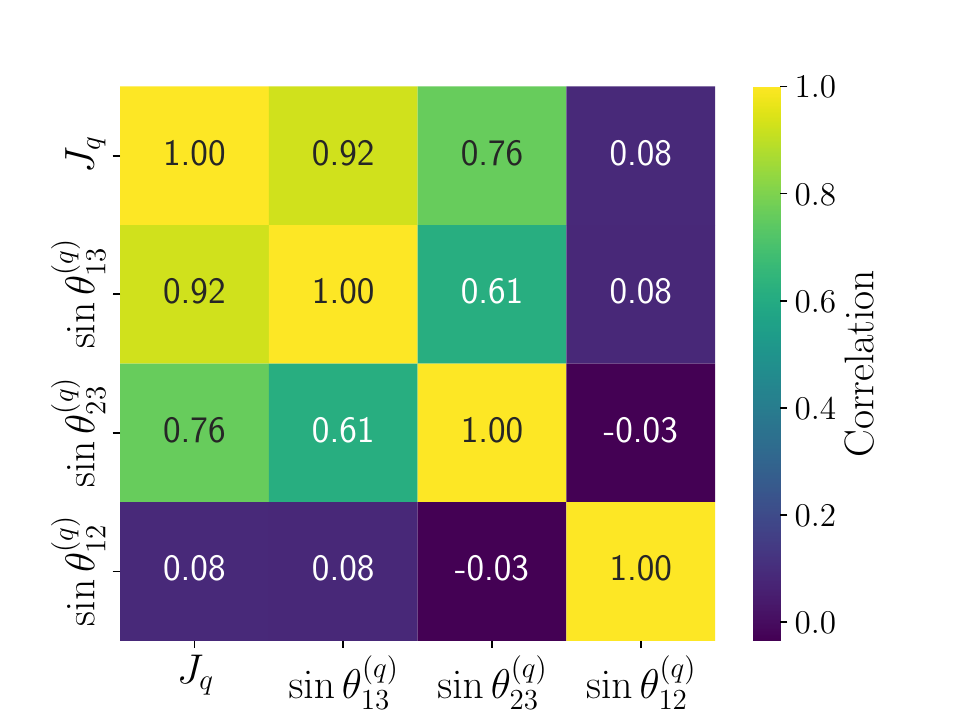}} 
\subfigure[]{\includegraphics[scale=.39]{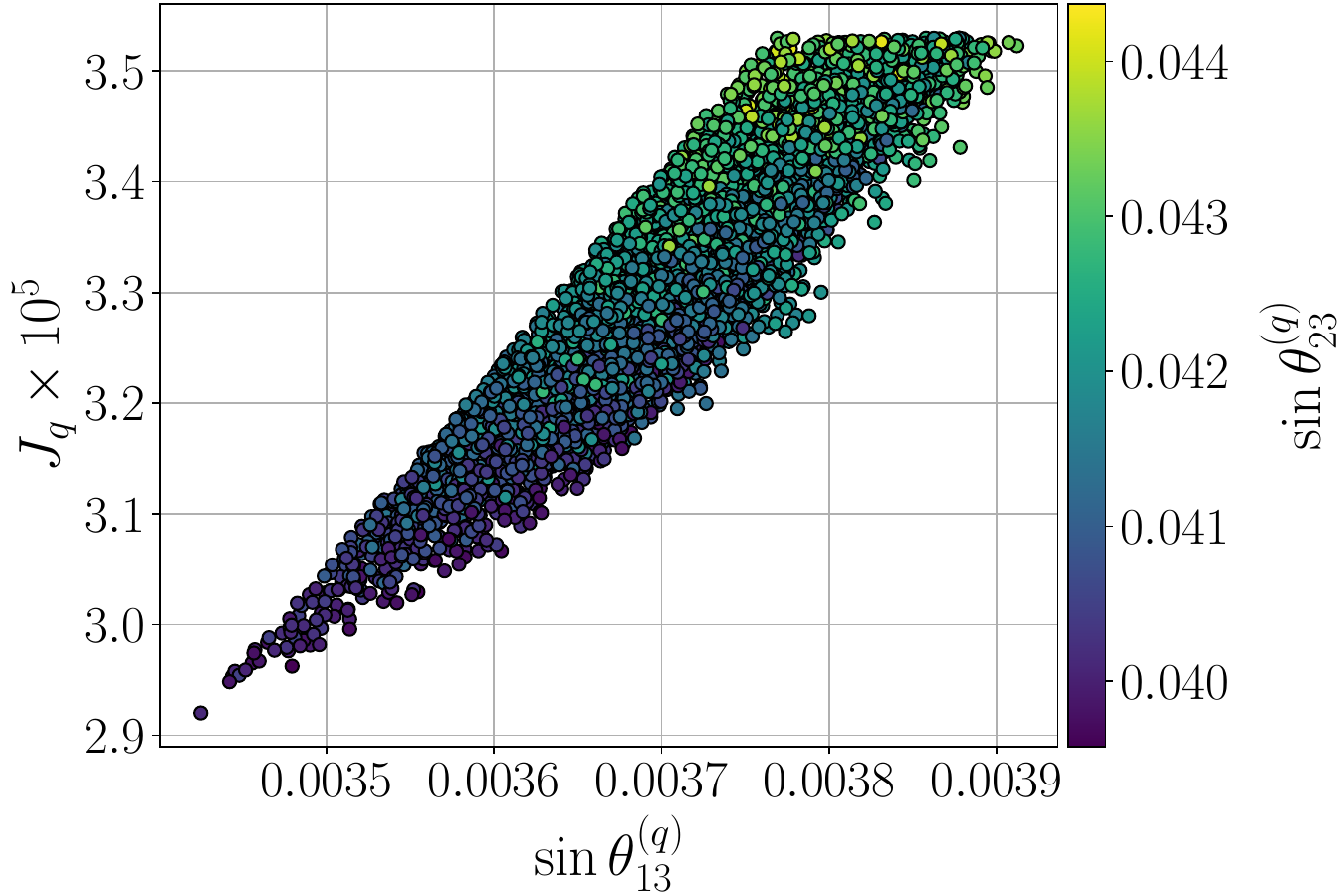}}
\newline
\caption{(a) Correlation heat-map: The correlation matrix between the quark mixing angle and the Jarlskog invariant $J_q$. \\ (b) Correlation plot: A color plot depicting the correlation between $\sin \theta_{13}$ and $J_q$, where the color intensity signifies the correlation strength in the $\sin \theta_{13}-\sin \theta_{23}$ plane.}
\label{fig:correlationquarksector}
\end{figure}

As shown in Table \ref{quarkobs}, our model successfully fit for the ten physical observables in the quark sector.
The symmetries inherent in our model enable us to successfully explain the SM quark mass spectrum and mixing parameters, with quark sector effective free parameters of order unity.
The correlation between the quark mixing angle $\sin \theta_{13}$ and the Jarlskog invariant is depicted in Figure \ref{fig:correlationquarksector}. In the left panel (a), we observe a heat-map for our model, revealing a strong correlation intensity in the $\sin\theta_{13}$-$J_q$ plane, with a correlation of $0.92$. This indicates that the Jarlskog invariant is highly sensitive to variation in $\theta_{13}$, aligning with the  experimental observation of its small value. Conversely, variations in $\theta_{12}$ have a negligible impact on the Jarlskog invariant. A significant correlation is also observed in the $\sin\theta_{23}$-$J_q$ plane. By analyzing this heat map, we can refine our parameter space scan for our model of interest. As previously noted, the right panel (b) graphically illustrates the correlation between $\sin\theta_{13}$ and the Jarlskog invariant $J_q$. In this figure, the color represents the correlation in the $\sin\theta_{13}$-$\sin\theta_{23}$ plane.

Finally, to close this section, we concisely discuss the collider signatures of the exotic quarks in our model. The exotic up type $J_1$ and down $J_n$ ($n=2,3$) quarks can be produced in pairs via the gluon fusion mechanism and Drell-Yan annihilation. Once produced, they can decay in several ways: $J_1\to H^+d_i$, $J_1\to W^{\prime +}d_i$, while for the $J_n$ quarks, the decay modes include $J_n\to H^-u_i$, $J_n\to W^{\prime -}u_i$ ($i=1,2,3$). Here, $H^{\pm}$ are physical electrically charged scalars associated with the third component of the $SU(3)_L$ $\rho$ scalar triplet. The electrically charged scalars $H^{-}$ ($H^{+}$) can further decay into the SM-charged lepton (antilepton) and neutrino. Consequently, the pair production of exotic quarks at a proton-proton collider can produce a final state composed of two jets, opposite sign dileptons, and missing energy. Thus, observing an excess of events involving the SM background in the opposite sign dileptons, two jets, and missing energy can be a possible signal of support of this model at the LHC. A detailed analysis of the collider signatures of exotic quarks in theories with extended $SU(3)_C\times SU(3)_L\times U(1)_X$ is provided in \cite{Cao:2016uur}. A comprehensive study of the exotic quark production and decay modes at colliders in our model goes beyond the scope of this work and is deferred for a future publication.

\begin{table}[]
\begin{center}
\begin{tabular}{c|l|l}
\toprule[0.2mm]
\hline
Observable & Model value & Experimental value \\ \hline \hline
$m_{u}(\text{MeV})$ & $\quad 2.33\pm 0.15$&  \quad $2.16_{-0.26}^{+0.49}$ \\ \hline
$m_{c}(\text{GeV})$ & \quad $1.26 \pm 0.06$&  \quad $1.27 \pm 0.02$ \\ \hline
$m_{t}(\text{GeV})$ & \quad $172.52 \pm 4.59$&  \quad $172.69\pm 0.30$ \\
\hline
$m_{d}(\text{MeV})$ & \quad $4.52 \pm 0.21$&  \quad $4.67_{-0.17}^{+0.48}$ \\ \hline
$m_{s}(\text{MeV})$ & \quad $95.5 \pm 5.6$&  \quad $93.4_{-3.4}^{+8.6}$ \\ \hline
$m_{b}(\text{GeV})$ & \quad $4.19\pm 0.15$&  \quad $4.18_{-0.02}^{+0.03}$ \\ \hline
$\sin \theta^{(q)} _{12}$ & \quad $0.225 \pm 0.005$&  \quad $0.22500 \pm 0.00067$ \\
\hline
$\sin \theta^{(q)} _{23}$ & \quad $0.0419 \pm 0.0024$&  \quad $0.04182_{-0.00074}^{+0.00085}$ \\
\hline
$\sin \theta^{(q)} _{13}$ & \quad $0.00372 \pm 0.00034$&  \quad $0.00369\pm 0.00011$ \\
\hline
$J_q$ & \quad $\left( 3.35 \pm 0.43\right) \times 10^{-5}$ &  \quad $\left(3.08^{+0.15}_{-0.13}\right) \times 10^{-5}$
\\ \hline
\bottomrule[0.2mm]
\end{tabular}
\end{center}
\caption{Model and experimental values of the quark masses and CKM parameters.}
\label{quarkobs}
\end{table}

\section{Lepton masses and mixings}
\label{leptons}

The charged lepton Yukawa interactions yield the following mass matrix for charged leptons:
\begin{eqnarray}
M_{\ell} &=&R_{\ell L}\text{diag}\left( m_{e},m_{\mu},m_{\tau}\right) ,\hspace{1cm}
\label{Ml} \\
R_{\ell L} &=&\frac{1}{\sqrt{3}}\left( 
\begin{array}{ccc}
1 & 1 & 1 \\ 
1 & \omega & \omega^{2} \\ 
1 & \omega^{2} & \omega%
\end{array}
\right) \left( 
\begin{array}{ccc}
\cos \alpha & \sin \alpha & 0 \\ 
-\sin \alpha & \cos \alpha & 0 \\ 
0 & 0 & 1%
\end{array}
\right) \left( 
\begin{array}{ccc}
\cos \gamma & 0 & -\omega \sin \gamma \\ 
0 & 1 & 0 \\ 
\omega^{2}\sin \gamma & 0 & \cos \gamma%
\end{array}
\right) ,\hspace{1cm}
\notag
\end{eqnarray}
where the charged lepton masses are:
\be
m_{e}=a_{1}^{\left( \ell\right) }\la  ^{8}\fr{v_{\rho }}{\sqrt{2}},\hspace{1cm}m_{\mu }=a_{2}^{\left(\ell\right) }\la  ^{5}\fr{v_{\rho }}{\sqrt{2}},
\hspace{1cm}m_{\tau }=a_{3}^{\left( \ell\right) }\la  ^{3}\fr{v_{\rho }}{\sqrt{2}}.  \label{leptonmasses}
\ee%
where $a_{i}^{\left( \ell \right) }=y_{i}^{(\ell)}$ ($i=1,2,3$) are $\mathcal{O}(1)$
parameters and we assumed that $y_{1,3}^{(\ell)}=z_{1,3}^{(\ell)}$. The neutrino
Yukawa interactions of Eq. (\ref{Lyl}) yield the following neutrino mass
terms:
\be
-\mathcal{L}_{mass}^{\left( \nu \right) }=\fr{1}{2}\left(
\begin{array}{ccc}
\overline{\left( \nu _L\right) ^{C}} & \overline{\nu _{R}} & \overline{%
N_{R}}%
\end{array}%
\right) M_{\nu }\left(
\begin{array}{c}
\nu _L \\
\left( \nu _{R}\right) ^{C} \\
\left( N_{R}\right) ^{C}%
\end{array}%
\right) +H.c,  \label{Lnu}
\ee%
where $\nu _{iR}\equiv \left( \nu _{iL}^{c}\right) ^{C}$, and the neutrino
mass matrix reads:
\be
M_{\nu }=%
\begin{pmatrix}
0_{3\times 3} & M_{\nu {D}} & 0_{3\times 3} \\
M_{\nu {D}}^{T} & 0_{3\times 3} & M_{\chi } \\
0_{3\times 3} & M_{\chi }^{T} & M_{R}%
\end{pmatrix}%
,  \label{eq_Mnu}
\ee%
and the sub matrices are given by:
\bea
M_{\nu {D}} &=&x_{\rho }\fr{v_{\rho }v_{\eta }v_{\si  {1}}^{2}}{2\La
^{3}}\left(
\begin{array}{ccc}
0 & 1 & -1 \\
-1 & 0 & 1 \\
1 & -1 & 0%
\end{array}%
\right) ,  \crn
M_{\chi } &=&y_{1\chi }^{\left( L\right) }\fr{v_{\chi }v_{S_{1}}}{\sqrt{2}%
\La  }\left(
\begin{array}{ccc}
1+x & 0 & 0 \\
0 & \om ^{2}+x\om & 0 \\
0 & 0 & \om +x\om ^{2}%
\end{array}%
\right) =\left(
\begin{array}{ccc}
m_{N_{1}} & 0 & 0 \\
0 & m_{N_{2}}e^{i\beta _{1}} & 0 \\
0 & 0 & m_{N_{3}}e^{i\beta _{2}}%
\end{array}%
\right) ,  \crn
M_{R} &=&\left(
\begin{array}{ccc}
y_{2N}v_{\va  } & y_{1N}v_{\xi } & -y_{1N}v_{\xi } \\
y_{1N}v_{\xi } & y_{2N}v_{\va  } & y_{1N}v_{\xi } \\
-y_{1N}v_{\xi } & y_{1N}v_{\xi } & y_{2N}v_{\va  }%
\end{array}%
\right) \fr{v_{\si_{3}}v_{\si_{1}}^{2}v_{\si  {2}}^{4}}{\La
^{7}}=\left(
\begin{array}{ccc}
\ka _{2} & \ka _{1} & -\ka _{1} \\
\ka _{1} & \ka _{2} & \ka _{1} \\
-\ka _{1} & \ka _{1} & \ka _{2}%
\end{array}%
\right) \mu ,\hspace{1cm}
x=\fr{{%
y_{1\chi }^{\left( L\right) }}v_{S_{1}}}{{y_{2\chi }^{\left( L\right) }}%
v_{S_{2}}},  \crn
\ka _{1} &=&y_{1N},\hspace{1cm}\hspace{1cm}\ka _{2}=y_{2N}\fr{%
v_{\va}}{v_{\xi }},\hspace{1cm}\hspace{1cm}\mu =\fr{v_{\si
_{3}}v_{\si  _{1}}^{2}v_{\si  _{2}}^{4}v_{\xi }}{\La  ^{7}}.
\label{MRmass1}\eea

The light active masses arise from an inverse seesaw (ISS) mechanism and the
physical neutrino mass matrices are:
\bea
M_{\nu }^{\left( 1\right) } &=&M_{\nu _{D}}\left( M_{\chi }^{T}\right)
^{-1}M_{R}M_{\chi }^{-1}M_{\nu _{D}}^{T},  \label{Mnu1} \\
M_{\nu }^{\left( 2\right) } &=&-\fr{1}{2}\left( M_{\chi }+M_{\chi
}^{T}\right) +\fr{1}{2}M_{R},\hspace{1cm}\hspace{1cm}M_{\nu }^{\left(
3\right) }=\fr{1}{2}\left( M_{\chi }+M_{\chi }^{T}\right) +\fr{1}{2}%
M_{R},  \label{Mnu2}
\eea%
where $M_{\nu }^{\left( 1\right) }$ is the light active neutrino mass matrix
whereas $M_{\nu }^{\left( 2\right) }$ and $M_{\nu }^{\left( 3\right) }$ are
the exotic Dirac neutrino mass matrices. The physical neutrino spectrum is
composed of three light active neutrinos and six pseudo-Dirac exotic
neutrinos, with masses $\sim \pm v_{\chi }\sim \mathcal{O}(10)$ TeV and a
small splitting $\sim \mu $.

Sterile neutrinos can be produced in pairs at the Large Hadron Collider (LHC) through Drell-Yan annihilation, mediated by a heavy $Z^{\prime}$ gauge boson. These sterile neutrinos mix with the light active neutrinos, enabling their decay into Standard Model (SM) particles. As a result, the final decay products include an SM-charged lepton and a $W$ gauge boson. Therefore, an excess of events in the dilepton final states, surpassing the SM background, could serve as a potential signature of this model at the LHC. Investigations into inverse seesaw neutrino signatures at colliders, as well as the production of heavy neutrinos at the LHC, have been conducted in, for example, in \cite{AguilarSaavedra:2012fu,Das:2012ii,Das:2012ze,BhupalDev:2012zg,Helo:2018rll,Pascoli:2018heg}. A thorough investigation of the collider implications of our model is beyond the scope of this work and will be explored in future research.

Thus, the mass matrix for light active neutrinos reads:
\bea
M_{\nu }^{\left( 1\right) } &=&\left(
\begin{array}{ccc}
\ka _{2}a^{2}-2\ka _{1}ab+\ka _{2}b^{2} & ab\ka _{1}-b\ka
_{1}-b^{2}\ka _{2}-a\ka _{1} & a\ka _{1}+b\ka _{1}-a^{2}\ka
_{2}+ab\ka _{1} \\
ab\ka _{1}-b\ka _{1}-b^{2}\ka _{2}-a\ka _{1} & \ka
_{2}b^{2}+2\ka _{1}b+\ka _{2} & a\ka _{1}-\ka _{2}-b\ka
_{1}-ab\ka _{1} \\
a\ka _{1}+b\ka _{1}-a^{2}\ka _{2}+ab\ka _{1} & a\ka
_{1}-\ka _{2}-b\ka _{1}-ab\ka _{1} & \ka _{2}a^{2}-2\ka
_{1}a+\ka _{2}%
\end{array}%
\right) m_{\nu }\allowbreak ,\hspace{1cm}  \crn
m_{\nu } &=&\fr{x_{\rho }^{2}v_{\rho }^{2}v_{\eta }^{2}v_{\si
_{1}}^{4}}{2\left( y_{\chi }^{\left( L\right) }\right)
^{2}\left( 1+x\right) ^{2}v_{\chi }^{2}v_{S_{1}}^{2}\La  ^{4}}\mu ,\hspace{1cm}a=\fr{%
1+x}{\om ^{2}+x\om },\hspace{1cm}b=\fr{1+x}{\om +x\om ^{2}}.
\eea
The light active neutrino mass matrix $M_{\nu }^{\left( 1\right) }$ is
diagonalized as follows:
\be
R_{\nu }^{T}M_{\nu }^{\left( 1\right) }R_{\nu }=\left(
\begin{array}{ccc}
0 & 0 & 0 \\
0 & m_{2} & 0 \\
0 & 0 & m_{3}%
\end{array}%
\right) ,\hspace{1cm}R_{\nu }=\left(
\begin{array}{ccc}
\fr{1}{\sqrt{3}} & -\fr{1}{\sqrt{2}} & \fr{1}{\sqrt{6}} \\
\fr{1}{\sqrt{3}} & 0 & \fr{-2}{\sqrt{6}} \\
\fr{1}{\sqrt{3}} & \fr{1}{\sqrt{2}} & \fr{1}{\sqrt{6}}%
\end{array}%
\right) \left(
\begin{array}{ccc}
1 & 0 & 0 \\
0 & \cos \theta & \sin \theta e^{-i\phi } \\
0 & -\sin \theta e^{i\phi } & \cos \theta%
\end{array}%
\right) ,
\ee

and the light active neutrino masses are:
\
\bea
m_{1} &=&0,\hspace{1cm} \\
m_{2} &=&\fr{1}{2}A+\fr{1}{2}B-\allowbreak \fr{1}{2}\sqrt{%
A^{2}-2AB+B^{2}+4C^{2}}, \\
m_{3} &=&\fr{1}{2}A+\fr{1}{2}B+\fr{1}{2}\sqrt{A^{2}-2AB+B^{2}+4C^{2}}.
\eea

where:
\begin{eqnarray*}
A &=&\left( 2\ka _{2}a^{2}-2\ka _{1}ab-2\ka _{1}a+\fr{1}{2}\ka
_{2}b^{2}-\ka _{1}b+\fr{1}{2}\ka _{2}\right) m_{\nu }\allowbreak , \\
B &=&\left( \fr{3}{2}\ka _{2}b^{2}+3\ka _{1}b+\fr{3}{2}\ka
_{2}\right) m_{\nu }\allowbreak , \\
C &=&\left[ -\fr{1}{2}\sqrt{3}\left( b-1\right) \left( \ka
_{2}-2a\ka _{1}+b\ka _{2}\right) \right] m_{\nu }\allowbreak
\end{eqnarray*}
and 
\begin{equation}
\tan\left(2 \theta\right) \equiv t_{2\theta} = \frac{\sqrt{3} (b-1) \left((b+1) \kappa _2-2 a \kappa_1\right)}{\kappa _2 \left(-2 a^2+b^2+1\right)+2 \kappa _1 ((a+2) b+a)}.
\end{equation}

Note that we consider only the normal neutrino mass hierarchy,
which is favored over the inverted hierarchy by more than  $3\si  $ . For the normal neutrino mass hierarchy, the PMNS leptonic
mixing matrix takes the form:

\be
U=R_{\ell L}^{\dagger }R_{\nu }=\left(
\begin{array}{ccc}
\cos\al \cos\gamma  & \fr{(\om +2)(\sin \al  \cos \gamma  +\sin \gamma )}{
\sqrt{6}} & \fr{\om (\sin \al  \cos \gamma  -\sin \gamma )}{\sqrt{2}}
\\
\sin \al  & -\fr{(\om +2)\cos \al  }{\sqrt{6}} & -\fr{\om \cos
\al  }{\sqrt{2}} \\
-\om \cos \al  \sin \gamma  & \fr{(\om -1)(\cos \gamma  -\sin
\al \sin \gamma )}{\sqrt{6}} & \fr{(\om +1)(\sin \al \sin\gamma  +\cos \gamma )}{%
\sqrt{2}}%
\end{array}%
\right) \left(
\begin{array}{ccc}
1 & 0 & 0 \\
0 & \cos \theta & \sin \theta e^{-i\phi } \\
0 & -\sin \theta e^{i\phi } & \cos \theta%
\end{array}%
\right) .
\label{hh1}\ee

The charged lepton masses, the neutrino mass squared splittings, i.e, $%
\Delta m_{21}^{2}$ and $\Delta m_{31}^{2}$, leptonic mixing angles $\theta
_{12}^{(\ell)}$, $\theta _{23}^{(\ell)}$, $\theta _{13}^{(\ell)}$ and the Dirac
leptonic CP violating phase are well reproduced for the following values of
the lepton sector parameters:
\begin{equation}\label{fit}
\begin{aligned}
a_{1}^{(\ell)} & \simeq 0.427, \quad a_{2}^{(\ell)} \simeq 1.026, \quad a_{3}^{(\ell)} \simeq 0.882 , \\
A & \simeq 31.65 \pm 3.12\ \text{meV}, \quad B\simeq   27.46 \pm 2.24 \ \text{meV}, \quad C \simeq 20.79 \pm 9.31 \ \text{meV} , \\
\theta & \simeq (-107.89 \pm 3.33)^{\circ}, \quad \phi \simeq (-29.83 \pm 2.07)^{\circ} , \\
\alpha & \simeq (-21.63 \pm 1.15)^{\circ}, \quad \gamma \simeq (-337.70 \pm 8.85)^{\circ} ,
\end{aligned}
\end{equation}
or 
\begin{equation}\label{eq:fit2}
\kappa_1 = -0.655001, \quad \kappa_2=0.726924, \quad x=-0.695083 + 0.718929 i.
\end{equation}

As shown in Table \ref{tab:neutrinos-NH}, our model is consistent with the
experimental data on lepton masses and mixing. It is worth mentioning that
our model allows to successfully reproduce the experimental values of the
six physical observables of the neutrino sector with only four effective
parameters. Analogously to the quark sector (see Eq.~\eqref{eq:chiquarkfit}), the fit of the leptonic sector is obtained by minimizing the function $\chi^2_\ell$, which reproduces the values in Table~\ref{tab:neutrinos-NH} and the best-fit point from Eqs. \eqref{fit} and \eqref{eq:fit2}.

\begin{table}[t]
{\footnotesize \ }
\par
\begin{center}
{\footnotesize \ \renewcommand{\arraystretch}{1}
\begin{tabular}{c|c||c|c|c|c}
\toprule[0.2mm]
\hline
\multirow{2}{*}{Observable} & \multirow{2}{*}{\parbox{7em}{Model\\ value} }
& \multicolumn{4}{|c}{Neutrino oscillation global fit values (NH)} \\
\cline{3-6}
&  & Best fit $\pm 1\si $ \cite{deSalas:2020pgw} & Best fit $\pm 1\si $
\cite{Esteban:2020cvm} & $3\si $ range \cite{deSalas:2020pgw} & $3\si $
range \cite{Esteban:2020cvm} \\ \hline\hline
$\Delta m_{21}^{2}$ [$10^{-5}$eV$^{2}$] & $7.51 \pm 0.60$ & $7.50_{-0.20}^{+0.22}$ & $7.42_{-0.20}^{+0.21}$ & $6.94-8.14$ & $6.82-8.04$ \\ \hline

$\Delta m_{31}^{2}$ [$10^{-3}$eV$^{2}$] & $2.55\pm 0.21$ & $2.55_{-0.03}^{+0.02}$& $2.517_{-0.028}^{+0.026}$ & $2.47-2.63$& $2.435-2.598$ \\ \hline

$\theta^{(\ell)}_{12}(^{\circ })$ & $33.4495 \pm 2.1648$ & $34.3\pm 1.0$ & $33.44_{-0.74}^{+0.77}$& $31.4-37.4$ & $31.27-35.86$ \\ \hline

$\theta^{(\ell)}_{13}(^{\circ })$ & $8.54137 \pm 0.26299$ & $8.58_{-0.15}^{+0.11}$ & $8.57\pm 0.12$& $8.13-8.92$& $8.20-8.93$ \\ \hline

$\theta^{(\ell)}_{23}(^{\circ })$ & $49.04 \pm 2.05$ & $48.79_{-1.25}^{+0.93}$ & $49.2_{-1.2}^{+0.9} $ & $41.20-51.33$& $40.1-51.7$ \\ \hline

$\delta_{\text{CP}}(^{\circ })$ & $-87.1246 \pm 7.7968$ & $216_{-25}^{+41}$ & $197_{-24}^{+27}$ & $128-359$& $120-369$ \\ \hline
\bottomrule[0.2mm]
\end{tabular}
{\normalsize \ }}
\end{center}
\par
{\footnotesize \ }
\caption{Model and experimental values of the neutrino mass-squared diferences, leptonic mixing angles, and the CP-violating phase. Experimental values are taken from Refs.~\protect\cite{deSalas:2020pgw,Esteban:2020cvm}.}
\label{tab:neutrinos-NH}
\end{table}

\begin{figure}[]
\centering
\subfigure[]{\includegraphics[scale=.45]{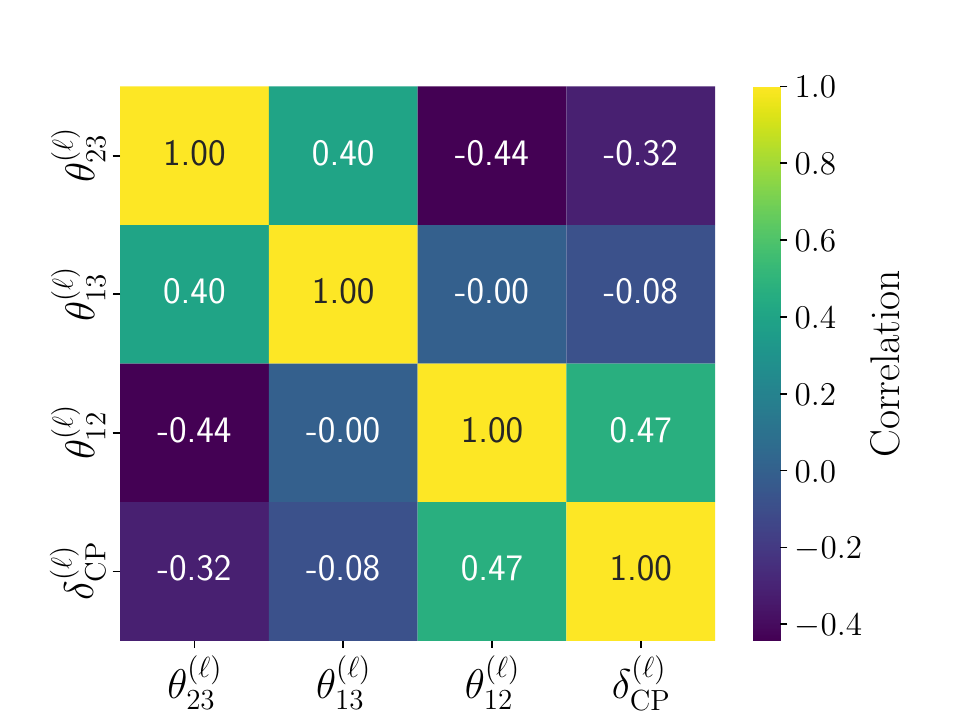}} \\
\subfigure[]{\includegraphics[scale=.38]{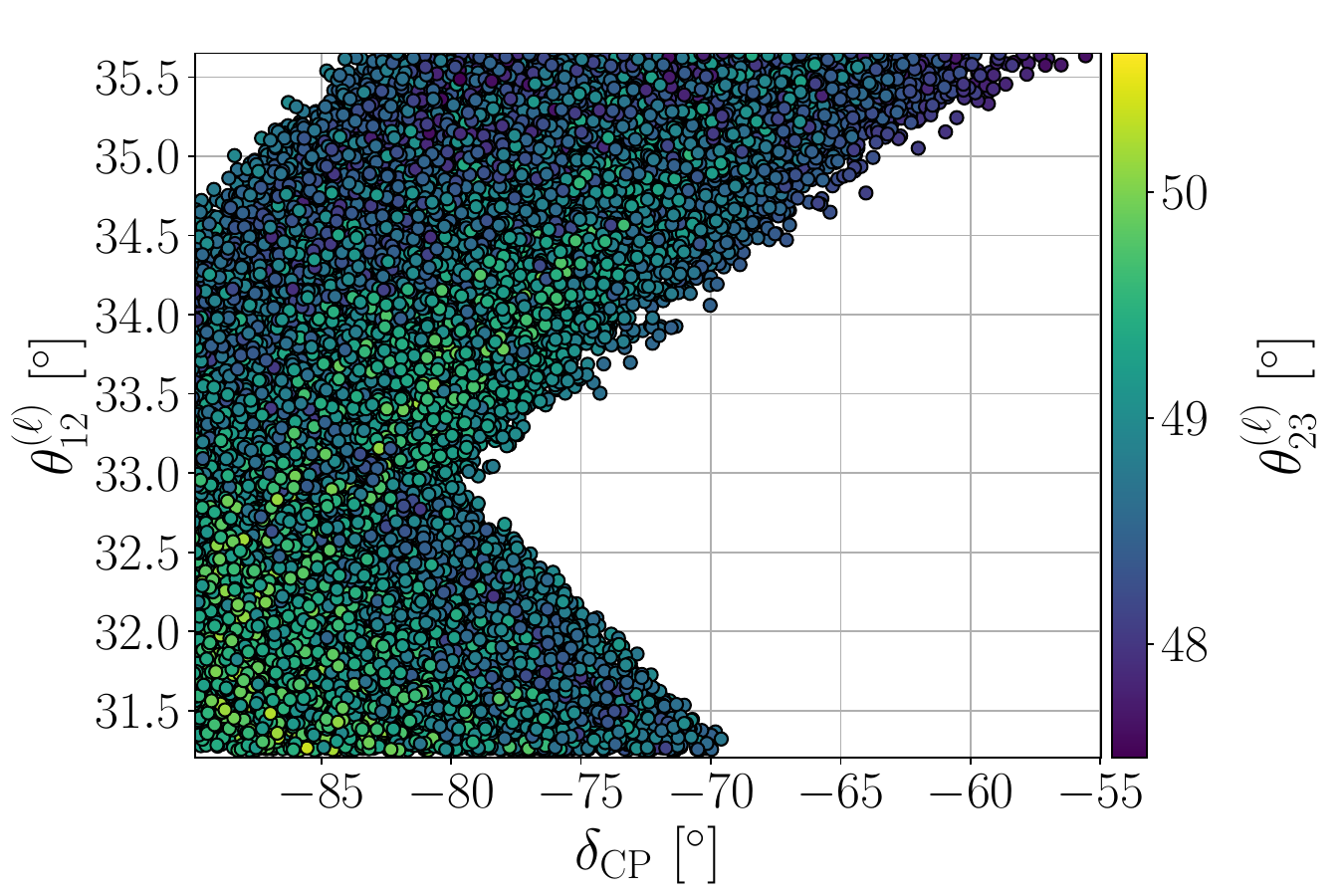}} \quad
\subfigure[]{\includegraphics[scale=.38]{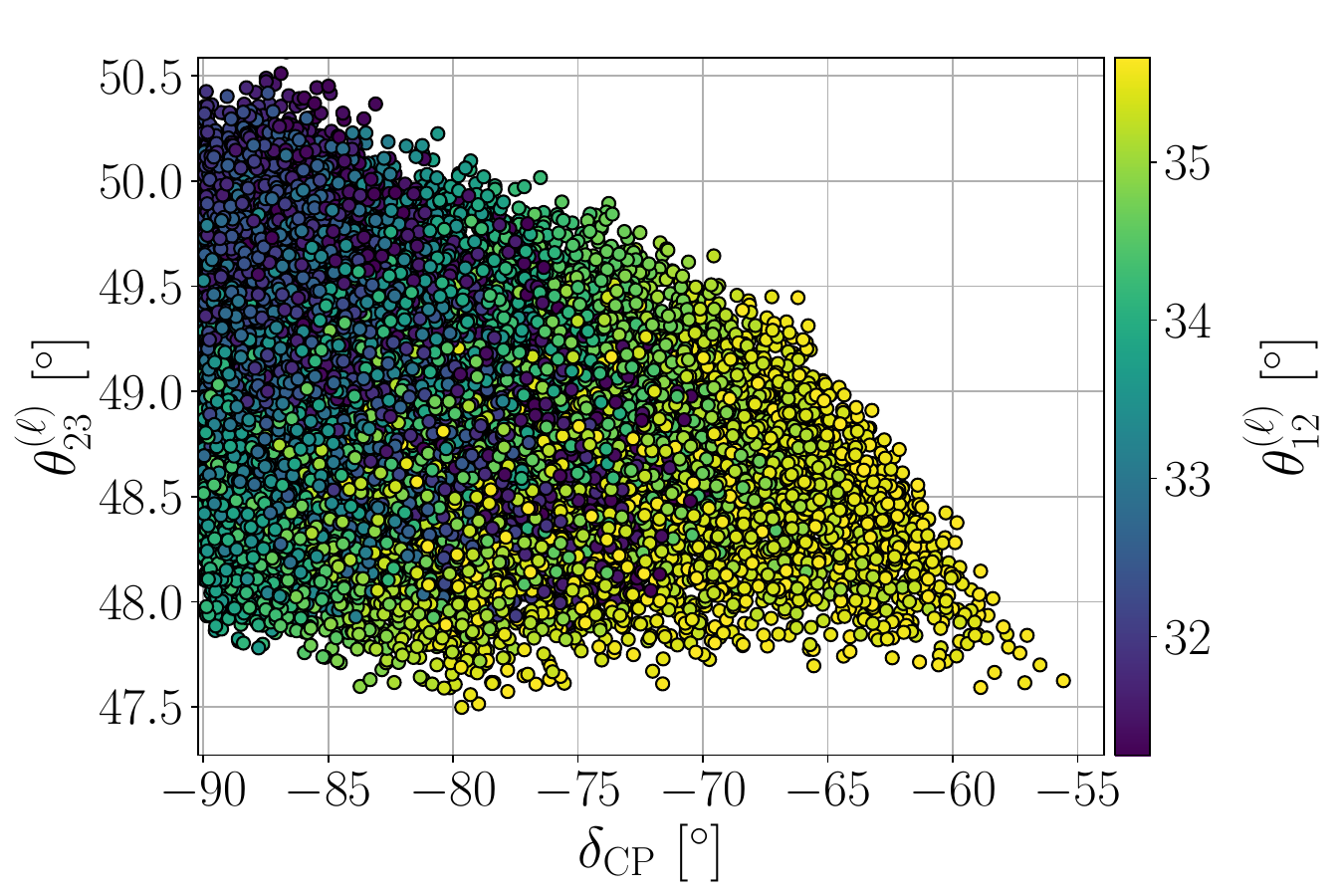}}
\newline
\caption{(a) Correlation heat-map: Illustrated the pairwise correlations between lepton mixing angle and the Dirac CP-violating phase. The intensity of color represents the strength of the correlation.  (b) Correlation between the solar mixing parameter and the leptonic Dirac CP phase, where the range of colors represents the correlation for the atmospheric mixing parameter our model. (c)Explored the correlation between the atmospheric mixing parameters and the Dirac CP-violating phase. }
\label{fig:correlationleptonsector}
\end{figure}

Figure \ref{fig:correlationleptonsector} illustrates the correlations among the lepton mixing parameters and the Dirac CP-violating phase. Panel (a) presents a heat-map, revealing the strongest correlations between the mixing angles $ \theta_{23}^{(\ell)} $ and $ \theta_{13}^{(\ell)} $. In contrast, the correlation with the phase $ \delta_{CP}^{(\ell)} $ is weaker, as indicated by the lower intensity in the values of the heat-map. Panel (b) displays the correlation between the solar mixing parameter $ \theta_{12}^{(\ell)} $ and the Dirac CP-violating phase $ \delta_{CP}^{(\ell)} $, which aligns with their $3\si $
experimentally allowed ranges.  Here the color range represents the correlation intensity for the atmospheric mixing parameter. Finally, panel (c) provides a similar visualization to panel (b), but focused on the atmospheric mixing parameter and the Dirac phase. This analysis is crucial for understanding the experimental constraints arising from the lepton mixing in our model.

Another observable predicted by our models is the effective Majorana neutrino mass parameter associated with neutrinoless double-beta decay $\left(0\nu\beta\beta\right)$. This process plays a pivotal role in investigating the Majorana nature of neutrinos, as it involves a nuclear decay that emits two electrons without any accompanying neutrinos. Its experimental observation would imply that neutrinos are Majorana fermions, violating the lepton number by two units and thereby pointing toward physics beyond the Standard Model.

The key parameter describing the amplitude of this process is the effective Majorana mass, $m_{ee}$, given by
\begin{eqnarray}
    m_{ee} = \biggl|\sum_{i} U_{ei}^{2}\,m_{\nu_i}\biggr|,
\end{eqnarray}
where $U_{ei}$ are the elements of the PMNS matrix and $m_{\nu_i}$ are the masses of the light active neutrinos. The theoretical estimation of this value depends on the neutrino mass hierarchy.

\begin{figure}[]
\centering
\includegraphics[width=0.5\linewidth]{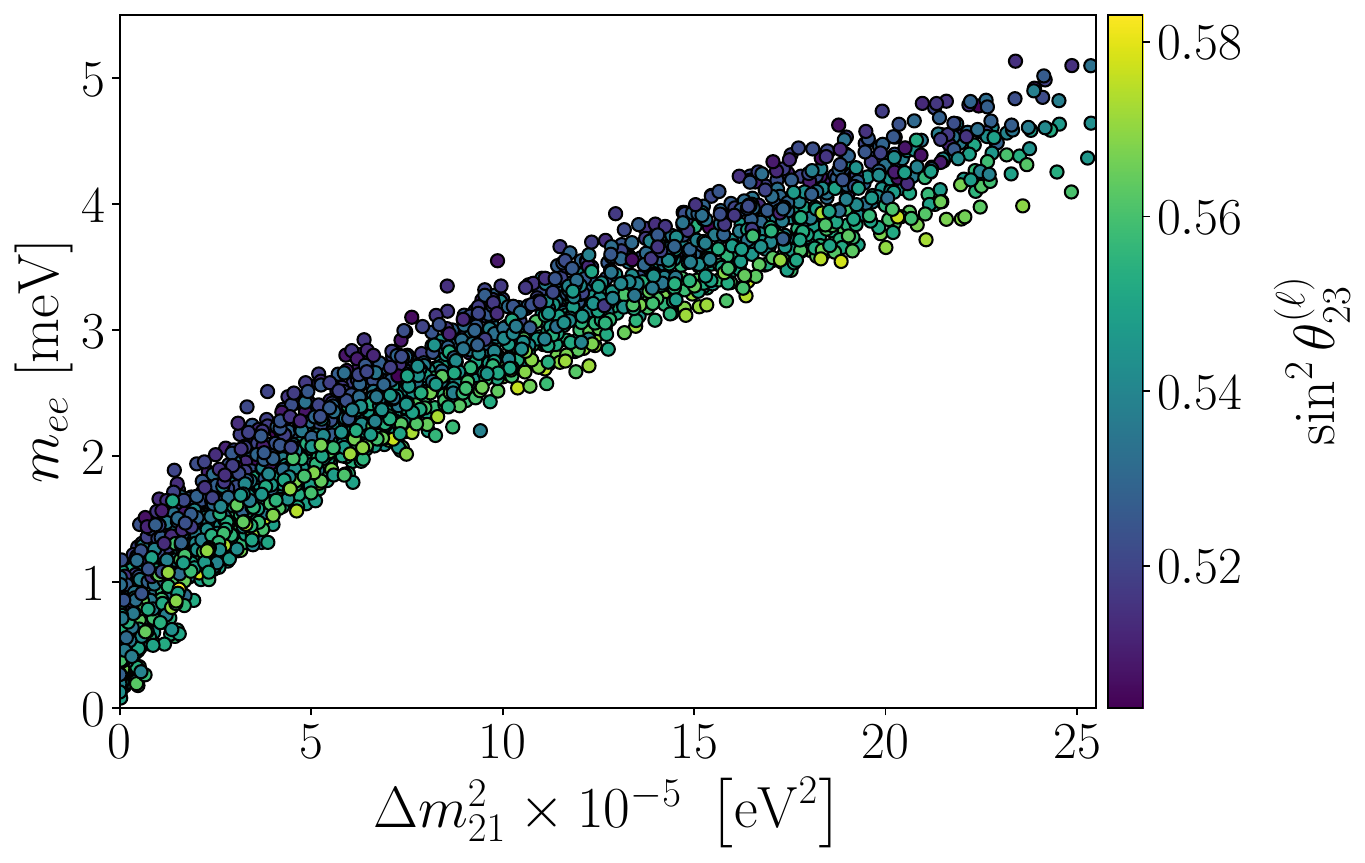}
\caption{Correlation between the squared neutrino mass difference $\Delta m_{21}^2$ and the effective Majorana neutrino mass parameter $m_{ee}$. The color scale represents the variation of the leptonic mixing parameter $\sin^2\theta_{23}^{(\ell)}$, highlighting its influence on the distribution of $m_{ee}$. 
}
\label{fig:mee}
\end{figure}

In Figure~\ref{fig:mee} illustrates the correlation between the effective Majorana neutrino mass parameter $m_{ee}$ and the neutrmass squared difference $\Delta m_{21}^2$,the color scale representing the leptonic mixing parameter $\sin^2\theta_{23}^{(\ell)}$. The predictions in this work are derived under the assumption of a normal neutrino mass hierarchy and are consistent with the experimentally allowed ranges for neutrino oscillation parameters. The effective Majorana neutrino mass parameter $m_{ee}$ is predicted to lie in the range $0.1\,\mathrm{meV} \lesssim m_{ee} \lesssim 5.0\,\mathrm{meV}$, which is well below the current experimental upper bound of $m_{ee} \lesssim 50\,\mathrm{meV}$ set by KamLAND-Zen \cite{KamLAND-Zen:2022tow} based on the non-observation of the decay of $0\nu\beta\beta$ for ${}^{136}\mathrm{Xe}$, with a half-life constraint of $T_{1/2}^{0\nu} > 2.0 \times 10^{26}\,\mathrm{yr}$ at the confidence level 90\%. Furthermore, our predictions are consistent with cosmological bounds from the Planck collaboration, which constrain the sum of neutrino masses to $\sum m_{\nu_i} < 0.115\,\mathrm{eV}$ at 95\% confidence level \cite{Planck:2018vyg}, taking into account our scenario of Eq.~\eqref{fit}, we obtain a value in the following range $0.01 \,\mathrm{eV} \lesssim \sum m_{\nu_i}  \lesssim 0.06\,\mathrm{eV} $ . The results presented here provide a robust framework to explore the parameter space of $m_{ee}$ within the context of normal hierarchy, highlighting the complementarity of neutrinoless double beta decay experiments and cosmological observations in probing the fundamental properties of neutrinos.

\section{Higgs and gauge bosons}
\label{scalarandgaugesector}

The model consists of $8$ electroweak gauge bosons $W^a_{\mu}$ of the $SU(3)_L$ gauge group and $X_{\mu}$ of $U(1)_X$. The corresponding covariant derivative is:
\be  \label{eq_Dmu}
D_{\mu}=\partial_{\mu} -ig T^aW^a_{\mu}-i g_X XT^9X_{\mu},
\ee
where $T^a$ are generators of the $SU(3)_L$ group, $T_9=I/\sqrt{6}$, $X$ is
the $U(1)_X$ charge, and
\be  \label{eq_gX}
\fr{g_X}{g}=\fr{3\sqrt{2}t_W}{\sqrt{3-t_W^2}},
\ee
with $t_W= \frac{\sin\theta_W}{\cos\theta_W}$, $\theta_W$ is  the Weinberg angle which is satisfied: $\sin \theta_W^2\simeq 0.231$%
.

After symmetry breaking, the model contains SM gauge bosons, including the massless photon $A_{\mu}$%
with their field given by
\begin{equation}
A_\mu = s_W W_{3\mu} -\frac{s_W}{\sqrt{3}} W_{8\mu} +\frac{\sqrt{1+2c_{2W}}}{\sqrt{3}} X_\mu,
\end{equation}
as well as two neutral weak gauge bosons 
\begin{equation}
 Z_\mu = c_W W_{3\mu}-t_W \left( -\frac{s_W}{\sqrt{3}} W_{8\mu} +\frac{\sqrt{1+2c_{2W}}}{\sqrt{3}} X_\mu\right), \hspace{0.5cm}   Z^\prime_\mu=\fr{\sqrt{1+2c_{2W}}}{\sqrt{3}s_W }W_{8 \mu} +\frac{t_W}{\sqrt{3}}X_\mu,
\end{equation}
with $c_W =\cos \theta_W, s_W=\sin \theta_W$. In the former
expressions, the neutral currents of $Z_\mu$
coincide with the weak neutral current of the SM. However, in the model considered, the physical fields
$Z_{1 \mu}, Z_{2 \mu}$ are defined
\be
Z_{1\mu} =c_z Z_\mu -s_z Z^\prime_\mu, \hspace{0.5 cm}
Z_{2 \mu} =s_z Z_\mu + c_z Z^\prime_\mu
\ee
where $s_{z}\equiv \sin\theta_z$, $c_{z}\equiv \cos\theta_z$, $c_{331}=\cos\theta_{331}$, $s_{331}=\sin\theta_{331}$ and
\be  \label{eq_t2z}
\tan 2\theta_z=-\fr{2 c_W \sqrt{3-t_W^2} v_{\rho }^2}{4 c_W^4
v_{\chi }^2+c_W^2 \left(t_W^2-3\right) v_{\rho }^2+v_{\rho }^2},\hspace{1cm} \tan\theta_{331}=-3\sqrt{2}\frac{g}{g_X}.
\ee

In the limit $\frac{v^2_{\rho}}{v^2_{\chi}}\sim O(\la ^6)\simeq 0$, we have 
\be
m^2_{Z_1}\simeq \frac{m^2_{W}}{c_W^2}, \hspace{0.5cm} m^2_{Z_2}\simeq \fr{g^2 v_{\chi }^2}{3-t_W^2}.
\ee

The single-charged gauge bosons $Z$ and $W^{\pm}$.
\be  \label{eq_Wboson}
W^{\pm}_{\mu}\equiv \fr{W^{1}_{\mu}\mp iW^{2}_{\mu}}{\sqrt{2}},\quad m_W=%
\fr{g v_{\rho}}{2}.
\ee

There are four new heavy-complex gauge bosons
\be  \label{eq_XYboson}
X^{0}_{\mu}\equiv \fr{W^{4}_{\mu}-iW^{5}_{\mu}}{\sqrt{2}}, \hspace{1cm}
Y^{\pm}_{\mu}\equiv \fr{W^{6}_{\mu} \mp iW^{7}_{\mu}}{\sqrt{2}},\ee
with their masses \be m_X =\fr{gv_{\chi}}{2}, \hspace{1cm} m^2_Y=m^2_X+m^2_W.
\ee

For user convenience, we provide the basis transformation matrix of the neutral gauge bosons, defined as
\begin{align}  \label{eq_Czzp}
\begin{pmatrix}
X_{\mu} \\
W^3_{\mu} \\
W^{8}_{\mu}%
\end{pmatrix}
=C_{zz^{\prime }}%
\begin{pmatrix}
A_{\mu} \\
Z_{1\mu} \\
Z_{2\mu}%
\end{pmatrix}%
, \quad C_{zz^{\prime }}=\left(
\begin{array}{ccc}
c_W s_{331} & c_{331} s_z-c_z s_{331} s_W & c_{331} c_z+s_{331} s_W s_z \\
s_W & c_W c_z & -c_W s_z \\
c_{331} c_W & -c_{331} c_z s_W-s_{331} s_z & c_{331} s_W s_z-c_z s_{331}\\
\end{array}
\right),
\end{align}

For the details of the Higgs potential, see Appendix~\ref{app_Higgs}. The model
contains a singly charged Higgs boson and a SM-like Higgs boson found by
LHC.

\section{FCNCs}
\label{FCNCs}
Since the quark generations are non-universal under the $SU(3)_L\times U(1)_X$ gauge symmetry, the corresponding tree-level flavor-changing neutral currents (FCNCs) must also be non-universal. Indeed, considering the interaction of neutral gauge bosons with fermions, we have:
\be
\mathcal{L}_{N.C}=-g \bar{f}\ga^\mu \left 
 \{T_3 W_{3 \mu}+ T_8 A_{8 \mu} +\frac{t}{\sqrt{6}}\left( Q-T_3 -\beta T_8\right)B_\mu \right\}f,
\ee
where $f$ runs over all fermion multiplets. The FCNCs associate only with $T_8$ for ordinary quarks, the revenant interactions are
\bea
\mathcal{L}_{N.C} &\supset & -g \bar{q}_{aL} \ga^\mu T_8 q_{aL} \left( A_{8 \mu}- \frac{\beta}{\sqrt{6}} t B_\mu\right) =-\frac{g \sqrt{3}}{\sqrt{3-t_W^2}} \bar{q}_{aL} \ga^\mu T_8 q_{aL}  Z_{\mu}^\prime \nonumber \\  &=&-\frac{g \sqrt{3}}{\sqrt{3-t_W^2}} \bar{q}_{L} \ga^\mu T_q q_{L}\left(-s_z Z_{1\mu}+c_z Z_{2 \mu} \right),
\eea 
with $T_q = \frac{1}{2\sqrt{3}} \text{Diag} \left( 1,-1,-1\right) $ for gauge states $q=u_L= \left(u_{1L}, u_{2L}, u_{3L} \right)^T$ or $q=d_L= \left(d_{1L}, d_{2L}, d_{3L} \right)^T$. The gauge states are related to the mass eigenstates by $ u_{L,(R)}=V_{u_L(u_R)}u^\prime_{L,(R)}$ and $ d_{L,(R)}=V_{d_L(u_R)}d^\prime_{L,(R)}$. 
The CKM matrix is given by $V_{\text{CKM}}=V_{u_L}^\dag V_{d_L}$. 
Due to the specific form of the quark mass matrix presented in Eq. (\ref{Mq}), the mixing matrix $ V_{d_L}$ is simply the identity matrix, specifically $V_{d_L} = \text{Diag} \left(1,1,1 \right)$. This indicates that $ V_{u_L}^\dag $ is equivalent to the CKM matrix, which means $V_{u_L}^\dag= V_{\text{CKM}}$. Consequently, transitioning to the mass eigenstates results in tree-level
FCNCs
only happen in $u$-quark sector
\bea
\mathcal{L}_{\text{FCNC}}&=& -\frac{g}{\sqrt{3-t_W^2}} \left\{ \left(V^{T}_{\text{CKM}} \right)_{1i}
\left(V_{\text{CKM}}^\dag \right)_{1j}  \right\} \bar{u}^\prime_{iL} \ga^\mu u_{jL} \left( -s_z Z_{1\mu}+c_z Z_{2 \mu}\right)\nonumber \\ &=& \vartheta_{ij}\bar{u}^\prime_{iL} \ga^\mu u_{jL} \left( -s_z Z_{1\mu}+c_z Z_{2 \mu}\right) , \label{FC2}\eea
with \bea \vartheta_{ij} =-\frac{g}{\sqrt{3-t_W^2}} \left\{ \left(V^{T}_{\text{CKM}} \right)_{1i}\left(V_{\text{CKM}}^\dag \right)_{1j}  \right\}. \label{FC2}\eea \\

Additional, the model also contains the scalar FCNCs associated to neutral scalars content in the triplet scalar Higgs. Two  physical states carrying both even $Z_2^{L}$ and CP charges, $(h_1, h_2)$, relates to two  scalar components $(\xi_\rho, \xi_\chi )$ by unitary rotation matrix as follows 
\bea \left( \xi_\rho, \xi_\chi, \right)^T =\mathcal{R} \left(h_1,h_2\right)^T.\eea
 
There is a tiny mixing between the two components $\xi_\rho$ and $\xi_\chi$ because $\xi_\rho$ is a doublet of the $SU(2)_L$ group, while $\xi_\chi$ is a singlet of this group. This implies that the matrix elements 
$\mathcal{R}_{11} \simeq \mathcal{R}_{22} \sim 1$ 
and $\mathcal{R}_{12} \simeq \mathcal{R}_{21} \sim \frac{v_\rho}{v_\chi} \sim \la^2$. 
After changing to the physical states, we obtain the scalar FCNCs associated with the CP-even scalar Higgs:
\bea \mathcal{L}_{\text{FCNC}}^{\text{scalar}}= \Ga_{\al \al^\prime}^{h_i} \bar{u}_{\al L} u_{\al^\prime R} h_i+ h.c.
\eea
with  \bea \Ga_{\al \al^\prime}^{h_i}=\frac{m_{u_\al^\prime}}{v_\rho}\left( \mathcal{R}_{1i}+ \mathcal{R}_{2i} \la^2-1\right) \left( V_{\text{CKM}}^T\right)_{1 \al}\left( V_{\text{CKM}}^\dag\right)_{1 \al^\prime} \simeq \frac{m_{u_\al^\prime}}{v_\rho} \la^4 \left( V_{\text{CKM}}^T\right)_{1 \al}\left( V_{\text{CKM}}^\dag\right)_{1 \al^\prime}. \label{FC3}\eea
The coefficients given by Eqs. (\ref{FC2}) and (\ref{FC3}) show that $\vartheta_{ij} \gg \Ga_{ij}$. The $\tan 2 \theta_z \simeq \la^4$. These mean that the FCNCs associated with the neutral scalars and $Z_{1\mu}$ are suppressed compared to those of the new neutral gauge bosons $Z_{2\mu}$. Hence, we may omit the FCNC contributions related to the neutral Higgs bosons and $Z_{1\mu}$in the following studies.\\

\begin{figure}[]
\centering
\includegraphics[width=0.4\linewidth]{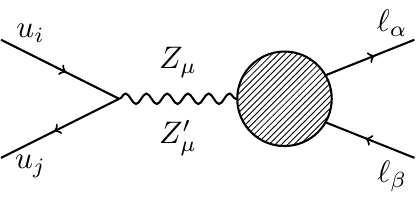}
\caption{Feynman diagram representing flavor-changing neutral currents (FCNC) in the up-quark sector. The vertex involves the neutral bosons $Z_\mu$ and $Z'_\mu$.}
\label{fig:fcnc}
\end{figure}

We want to emphasize that, unlike previous studies, our model predicts the existence of tree-level FCNCs primarily associated with the $u$-quark. 
This feature is shown in Figure~\ref{fig:fcnc}, where the Feynman diagram depicts FCNCs in the up-quark sector, mediated by the neutral bosons $Z_{\mu}$ and $Z^{\prime}_{\mu}$. 
Consequently, our model is not tightly constrained by B-meson oscillations but only by D-meson oscillations. At tree level, the effective Lagrangian for the D-meson oscillation process is as follows:
\bea
\mathcal{L}_{Z^\prime \text{eff}}^{D_0-\bar{D}_0} = \mathcal{G}^\prime \frac{m_{Z_1}^2}{m_{Z_2}^2} \left | \left( V_{\text{CKM}}\right)_{11} \left( V_{\text{CKM}}^*\right)_{21}\right |^2 \left|\bar{u}^\prime_{1L}\ga^\mu u_{2L}^\prime \right|^2,
\eea
where $\mathcal{G}^\prime = \frac{4\sqrt{2}G_F}{1+4c_W^2}$, with $G_F$ being the Fermi constant. This effective Lagrangian contributes to the mass splittings $\Delta m_D$ between neutral mesons $D_0$ and $\bar{D}_0$ as follows:
\bea
\Delta m_{D} = \mathcal{G}^\prime \frac{m_{Z_1}^2}{m_{Z_2}^2}\left| \left( V_{\text{CKM}}\right)_{11}\left( V_{\text{CKM}}^*\right)_{21}\right|^2 f_D^2 B_D \eta_D m_D.
\eea
Values for the bag parameter $B_D$, decay constant $f_D$, QCD correction factor $\eta_D$, and the D-meson mass are taken from \cite{PhysRevD.107.063005}: 
\bea
\sqrt{B_D}f_D =187 \text{MeV}, \hspace{1cm} \eta_D =0.57, \hspace{1cm} m_D =(1865\pm 0.0005) \text{MeV}.
\eea
The SM predicted value for $\Delta m_D$ is
\bea
\left( \Delta m_D \right)_{\text{SM}} = 10^{-14} \text{MeV},
 \eea
 while the experimental value is
 \bea
\left( \Delta m_D \right)_{\text{exp}} =\left(6.25316^{+2.69873}_{-2.8962}\right) \times 10^{-12} \text{MeV}.
 \eea

 This means that the main contribution to the D-meson mass difference comes from the NP. From the experimental bound, we obtain the lower bound on
 the $Z_2$-boson mass: $m_{Z_2}> 173 \text{TeV}$. \\

 We would like to point out that if we organize the third quark family transform differently from the first two families under the symmetry group $SU(3)_c \times U(1)_L$, the NP contribution to the D-meson mass difference becomes proportional to $\left|\left( V_{\text{CKM}}\right)_{13}\left( V_{\text{CKM}}^*\right)_{32} \right|^2$ 
 instead of being proportional to $\left| \left( V_{\text{CKM}}\right)_{11}\left( V_{\text{CKM}}^*\right)_{21}\right|^2$, as considered in the version considered. In this scenario, the constraint on the oscillation of $D_0$ and $\bar{D}_0$ results in a lower bound on the mass of the new neutral boson, which is approximately a few TeV.

Since the three lepton families transform identically under  $SU(3)_c \times U(1)_L$, lepton FCNCs are forbidden at the tree level but occur at the loop level, as demonstrated in \cite{Hong:2024swk}. The model allows for lepton flavor violation (LFV) via one-loop penguin diagrams in meson decays, particularly in D-meson decays. The general four-fermion Hamiltonian describing the interaction for the process  $u^\prime_1 \to u^\prime_2 l_\al^+ l_\beta^-$ can be expressed as follows: 
 \bea
 \mathcal{H}_{\text{eff}} =\frac{4 G_\text{F}}{\sqrt{2}} \left( C_9 O_9+C_{10} O_{10}\right),
 \eea
where  \bea C_9 &\simeq& \frac{2}{1+4c_W^2} \frac{4 \pi^2}{e^2}\left( \frac{m_{Z_1}^2}{m_{Z_2}^2}\right)\left( V_{\text{CKM}}\right)_{11}\left( V_{\text{CKM}}^*\right)_{12} \left( \bar{a}_l +\bar{a}_r \right), \nonumber \\ C_{10} &\simeq& \frac{2}{1+4c_W^2}\frac{4 \pi^2}{e^2}\left( \frac{m_{Z_1}^2}{m_{Z_2}^2}\right)\left( V_{\text{CKM}}\right)_{11}\left( V_{\text{CKM}}^*\right)_{12} \left( \bar{a}_l -\bar{a}_r \right). \eea
In our case, the expressions for $\bar{a}_{l(r)}$  are determined similarly to those for $a_{l(r)}$ given in \cite{Hong:2024swk}. In that work, they predicted $\text{Br}\left(Z \to \mu e \right) \sim 10 \times 10^{-12}$, which resulted in coefficient values of $a_{l(r)} \simeq 10^{-3}$. Combining this with the limit  $m_{Z_2} > 173 \text{TeV}$, we can roughly estimate the Wilson coefficients $C_{10}, C_9 \sim 10^{-10}$. Due to these small values, we predict suppressed values for the branching ratios of charged lepton flavor violation in D-meson decays, well below the experimental upper bound reported in  \cite{LHCb:2015pce}. \\
 Due to the light-heavy neutrino mixing matrix elements, the model also yields the box diagrams for the transition $ q_i \to q_j l_\al^+ l_\beta^-$. The internal lines can be either a light neutrino or heavy neutrino, and either the SM $W_\mu^{\pm}$ or new charged gauge bosons $Y_{\mu}^\pm$. However, the contributions to the Wilson coefficients from the box diagrams are very small\cite{Awasthi:2024nvi, Gagyi-Palffy:1994xuz}. Hence, we do not consider them in our work.

\section{Lepton flavor violating decays $e_j\rightarrow e_i$
\label{LFV}}

Based on the definition of the covariant derivative associated with the electroweak gauge group $SU(3)_L \times U(1)_X$, the charged gauge boson component $V_\mu^{CC}$ of the  derivative $D_\mu$ associated in the model under consideration is given by:
\begin{equation}
V_\mu^{C C} \equiv \frac{1}{\sqrt{2}}\left(\begin{array}{ccc}
0 & W_\mu^{+} & 0 \\
W_\mu^{-} & 0 & Y_\mu^{-} \\
0 & Y_\mu^{+} & 0
\end{array}\right).
\end{equation}
Given that the one-loop contributions from neutral Higgs bosons, $Z$ and $Z^{\prime }$
bosons involve couplings only to the SM charged lepton, we will be ignored them in this work. Therefore, we will focus solely on the couplings of charged
leptons to both heavy neutral leptons and charged gauged bosons, which are
relevant for one-loop contributions to LFV decays.
The
calculation will be performed on the unitary gauge.

Left-handed charged leptons are transformed from the flavor basis $l_{iL,R}$ to the
physical basis $e_{iL,R}$ using the matrix $R_{\ell L}$ from Eq.~\eqref{Ml} 
$l_{iL} = R_{\ell L,ij} e_{jL} $ 
with $i,j=1,2,3$.
The neutrino mass matrix $M_\nu$ in Eq. \eqref{eq_Mnu} is diagonalized by the unitary 
matrix $U_\nu$:
\begin{align}  \label{define_Unu}
U_\nu^T M_\nu U_\nu &= \hat{M}_\nu = \begin{pmatrix} 
\hat{m}_\nu & \textbf{0}_{3\times 9} \\
\textbf{0}_{9\times 3} & \hat{m}_N 
\end{pmatrix},
\end{align}
where $\hat{m}_\nu=\mathrm{diag}(m_{n_1},\;m_{n_2},\;m_{n_3})$ and $\hat{m}_N =\mathrm{diag}(m_{n_4},\;m_{n_5},...,\;m_{n_9})$ are the diagonal masses matrices for the light and heavy Majorana neutrinos  $n_L=(n_{1L},n_{2L},...,n_{9L})$, respectively. 
This implies the following relationship between flavor and physical states:$(\overline{\nu^C_L} \; \overline{\nu_R} \; \overline{N_R})=\overline{n_R}U^T_\nu$ and $(\nu_L \; \nu^C_R \; N^C_R)^T= U_\nu n_L$. Due to the ISS
condition $|m_{D}|\ll |M_{\chi}|$, the light neutrino mixing matrix $R_\nu$
from Eq.~(\ref{hh1}) is approximately equal to $U_\nu$:  $R_{\nu,ij}\simeq U_{\nu,ij}$ for all $i,j=1,2,3$ .

The rotation matrix $U_\nu$ from Eq. \eqref{define_Unu} is given by \cite{Catano:2012kw}:
\begin{equation}
U_\nu=\left(\begin{array}{ccc}
R_\nu   & R_1 R_M^{(2)}     & R_2 R_M^{(3)} \\
-\frac{\left(R_1^{\dagger}+R_2^{\dagger}\right)}{\sqrt{2}} R_\nu & \frac{(1-S)}{\sqrt{2})} R_M^{(2)} & \frac{(1+S)}{\sqrt{2}} R_M^{(3)} \\
-\frac{\left(R_1^{\dagger}-R_2^{\dagger}\right)}{\sqrt{2}} R_\nu & \frac{(-1-S)}{\sqrt{2}} R_M^{(2)} & \frac{(1-S)}{\sqrt{2}} R_M^{(3)}
\end{array}\right),
\end{equation}
where the submatrices $R_1$, $R_2$ and $S$ are defined from the Eq.~\eqref{MRmass1} as follows:
\begin{equation}
R_1 \simeq R_2 \simeq \frac{1}{\sqrt{2}} M_{\nu D}^* M_{\chi}^{-1}, \quad S=-\frac{1}{4} M_{\chi}^{-1} M_R,
\label{nuNmixingangles}
\end{equation}
and $R_M^{(2)}$, $R_M^{(3)}$ diagonalize the exotic neutrino matrices.

The covariant kinetic terms of $L_{iL}$ give the following couplings of charged
gauge bosons with leptons:
\begin{align}  \label{eq_LeeV}
\mathcal{L}_{V ll}&= g\sum_{i=1}^3\overline{L_{iL}}V^{CC}_{\mu}\ga ^{%
\mu}L_{iL}=\fr{g}{\sqrt{2}} \sum_{i=1}^3\left( \overline{\nu_{iL}}%
\ga ^{\mu}l_{iL} W^+_{\mu} + \overline{(\nu_{iR})^C}\ga ^{\mu}l_{iL}
Y^+_{\mu}\right) +H.c.  \crn
&=\fr{g}{\sqrt{2}}\sum_{a=1}^9\sum_{i,j=1}^3\left(\overline{n_{a}}%
\ga ^{\mu}U^{\dagger}_{\nu,aj}R_{\ell L,ji} P_L e_{i}W^+_{\mu} + \overline{n_{a}%
}\ga ^{\mu}U^{\dagger}_{\nu,a(j+3)}R_{\ell L,ij}P_Le_{i} Y^+_{\mu} \right) +H.c.
\end{align}
with $P_{L}=\left(1 - \gamma_5\right) / 2.$ \\

Processes like $\mu \rightarrow e \gamma$ are crucial for probing beyond SM physics, including 3-3-1 models. Based on the developments presented in~\cite{Hue:2017lak,Hue:2017lak,Nguyen:2018rlb}., we explore the LFV processes arising from non-universal couplings of the standard charged leptons to the gauge bosons. In our model, LFV is directly linked to the masses and interactions of exotic neutrinos and their associated gauge bosons. Experimental limits, particularly MEG ones, impose stringent constraints on the parameter space of 3-3-1 models. Future experiments, conversion in aluminum and titanium nuclei, will further refine these constraints and provide insights into the scale of new physics.

\begin{figure}[]
    \centering
\subfigure[]{\includegraphics[width=0.49\linewidth]{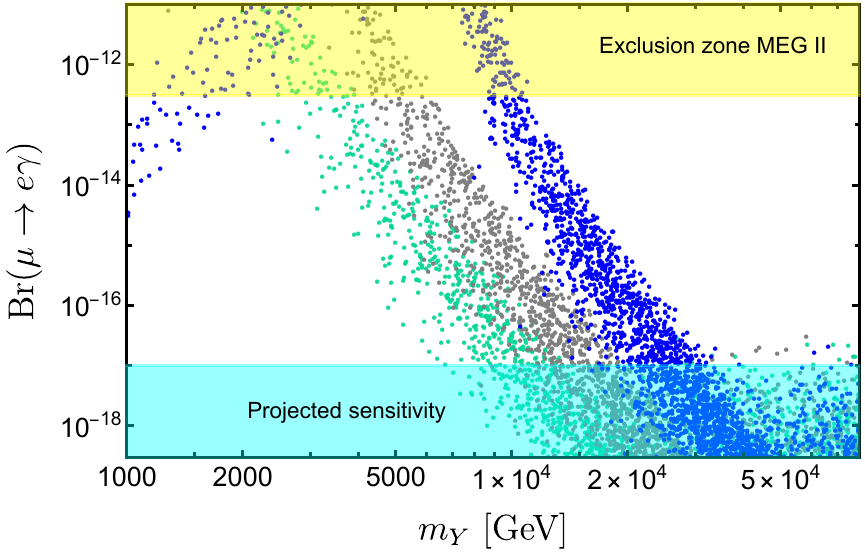}}
\subfigure[]{\includegraphics[width=0.49\linewidth]{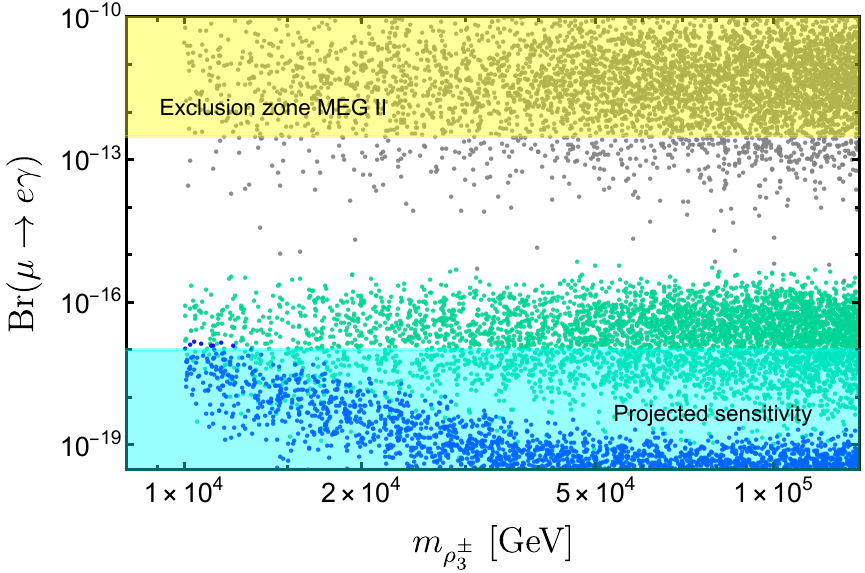}}
\caption{Branching ratio for $\mu \rightarrow e\gamma$ as a function of the mass of the gauge boson $Y$ (panel a) and the charged scalar $\rho_3^\pm$ (panel b). The upper shaded region corresponds to an area excluded by the MEG II experiment~\cite{MEGII:2023ltw} and the lower shaded region corresponds to the expected sensitivities of the next generation of experiments using aluminium as targets~\cite{Bernstein:2013hba}.}
\label{fig:lfv}
\end{figure}

The analytic expression for the branching ratio (Br) for the decay of $e_i \rightarrow e_j \gamma$ ($m_{e_i}>m_{e_j}$) is given by \cite{Hue:2017lak, Ma:2001mr, Toma:2013zsa, Vicente:2014wga, Lindner:2016bgg}:
\begin{align}  \label{breijga}
\mathrm{Br}(e_i\rightarrow e_j\ga)&= \fr{12\pi^2}{G_F^2}\left|\Gamma_{ij}\right|^2\times \mathrm{Br}(e_i\rightarrow e_j\overline{\nu}_j \nu_i),
\end{align}
where $\Gamma_{ij}$ represents the one-loop contribution from virtual charged gauge bosons and Majorana neutrinos in the loop diagrams. The contribution can be expressed as: $\Gamma_{ij}=\Gamma^{W}_{ij} +\Gamma^{Y}_{ij}+\Gamma^{\rho_3^{\pm}}_{ij}$, where:
\begin{equation}
\begin{aligned}
\Gamma^W_{ij}&=-\fr{eg^2}{32\pi^2m_W^2} \sum_{k=1}^9%
\sum_{a,b=1}^3(R_{lL}^*)_{ib}
(U^*_\nu)_{bk}(R_{lL})_{ja}(U_\nu)_{ak}\mathcal{F}\left(\fr{m^2_{n_k}}{m_W^2} \right),  \\
\Gamma^Y_{ij}&= -\fr{eg^2}{32\pi^2m_Y^2} \sum_{k=1}^9 \sum_{a,b=1}^3 (R_{lL}^*)_{ib} (U^*_\nu)_{(b+3)k}(R_{lL})_{ja}(U_\nu)_{(a+3)k}\mathcal{F}\left( \fr{m^2_{n_k}}{m_Y^2}\right),  \\
\Gamma^{\rho_3^{\pm}}_{ij}&=\frac{e^2}{2}\sum_{k=1}^9 \frac{z_{kj}^*\, z_{ki}}{(4 \pi)^2} \frac{1}{m_{\rho_3^{\pm}}^2}\mathcal{G}\left(\fr{m^2_{n_k}}{m_{\rho_3^{\pm}}}\right)
\end{aligned}
\end{equation}
with
\begin{equation}
z_{ki} =\sum_{r=1}^3 \sum_{s=1}^3 y_{\rho_3^-}^{rs} \left(R_{lL}^\dagger\right)_{ir}\left(\tilde{U}_{\nu}\right)_{sk},
\end{equation}
and the loop functions $\mathcal{F}(z)$ and $\mathcal{G}(z)$ are defined as:
\begin{eqnarray}
\mathcal{F}(z)&=&2(z+2) \mathcal{I}_3(z)-2(2 z-1) \mathcal{I}_2(z)+2 z \mathcal{I}_1(z)+1,    \notag\\
\mathcal{G}(z)&=&\frac{1 - 6 z + 3 z^2 + 2 z^3 - 6 z^2 \log z}{6 (1-z)^4}\,,
\end{eqnarray}
\begin{equation}
\mathcal{I}_n(z) = \frac{1}{4}\int_{0}^1 \frac{t^n}{z(t-1)-t} dt.
\end{equation}

An additional LFV process is the $\mu \to e$ conversion in nuclei. The conversion rate is defined as the ratio \cite{Lindner:2016bgg,CarcamoHernandez:2024ycd}
\begin{equation}
\text{CR}(\mu-e)=\frac{\Gamma\left(\mu^{-}+\operatorname{Nucleus}(A, Z) \rightarrow e^{-}+\operatorname{Nucleus}(A, Z)\right)}{\Gamma\left(\mu^{-}+\operatorname{Nucleus}(A, Z) \rightarrow \nu_\mu+\operatorname{Nucleus}(A, Z-1)\right)},
\end{equation}
where $A$ is the mass number and $Z$ is the atomic number of the nucleus.

In 331 models, the $\mu$-$e$ conversion is more sensitive than the decay $\mu \rightarrow e \gamma$, with the sensitivity restricted by approximately two orders of magnitude. Future experiments are expected to significantly improve these sensitivities, reaching an estimated limit of $10^{-17}$ for experiments using aluminum.
Taking the above into account, the estimation of the $\mu$-$e$ conversion can be made using the following relation \cite{Lindner:2016bgg}, which approximates the connection between the conversion 
 in aluminum nuclei and the decay rate $\mu \rightarrow e \gamma$:
\begin{equation} \label{eq:aproxBound}
\mathrm{CR}(\mu^- \mathrm{Al} \rightarrow e^- \mathrm{Al}) \approx \frac{1}{350} \mathrm{Br}(\mu \rightarrow e \gamma).
\end{equation}

The dominance of the photon in the $\mu$-$e$ conversion guarantees the applicability of the previous relation; this is fulfilled in our case, because at tree level we do not have neutral scalars that change flavor.

Figure~\ref{fig:lfv} displays the model predictions for the branching ratio $\text{Br}(\mu \rightarrow e \gamma)$ as a function of the exotic gauge boson mass $m_Y$ and the charged scalar mass $m_{\rho_3^\pm}$. In panel (a) the color-coded points represent different exotic neutrino mass values: green for  $1$ TeV,  gray for $2$ TeV, and blue for $5$ TeV, whereas in panel (b) the color-coded point represent different mass ranges of the exotic gauge boson $Y$.
The yellow shaded region, constrained by the MEG II experiment's limit of  $\text{Br}(\mu \rightarrow e \gamma) < 3.1 \times 10^{-13}$ \cite{MEGII:2023ltw}, excludes certain parameter space regions. 
The cyan shaded region indicates the projected sensitivity of future experiments like $\mu-e$ conversion in titanium and aluminum nuclei, with a sensitivity of $\text{CR}(\mu^{-} \mathrm{Al} \rightarrow e^{-} \mathrm{Al}) \lesssim 10^{-17}$ \cite{Bernstein:2013hba}. Numerically, the Yukawa couplings $y_{\rho_3^-}$ arising from the Dirac block fluctuate within the range $0.01$ to $0.1$, and the relevant couplings are given by $x_\rho = 1.99$ and $y_{1\chi }^{\left( L\right) } = 1.92$, which directly affect the magnitude of the contributions from gauge bosons and exotic neutrinos to LFV processes, since they control the active-sterile neutrino mixing angles given by Eq. (\ref{nuNmixingangles}), which determine the couplings of the neutrinos with SM charged leptons and charged gauge bosons, then giving rise to charged lepton flavor violating processes.

The most stringent limits for LFV arise from muon decay measurements, particularly from $ \mu \rightarrow e \gamma $. The latest experimental results impose a strict upper limit on the branching ratio of this process, represented in the upper part of the plot in Figure~\ref{fig:lfv}. 

\begin{figure}
\centering
\subfigure[]{\includegraphics[width=0.49\linewidth]{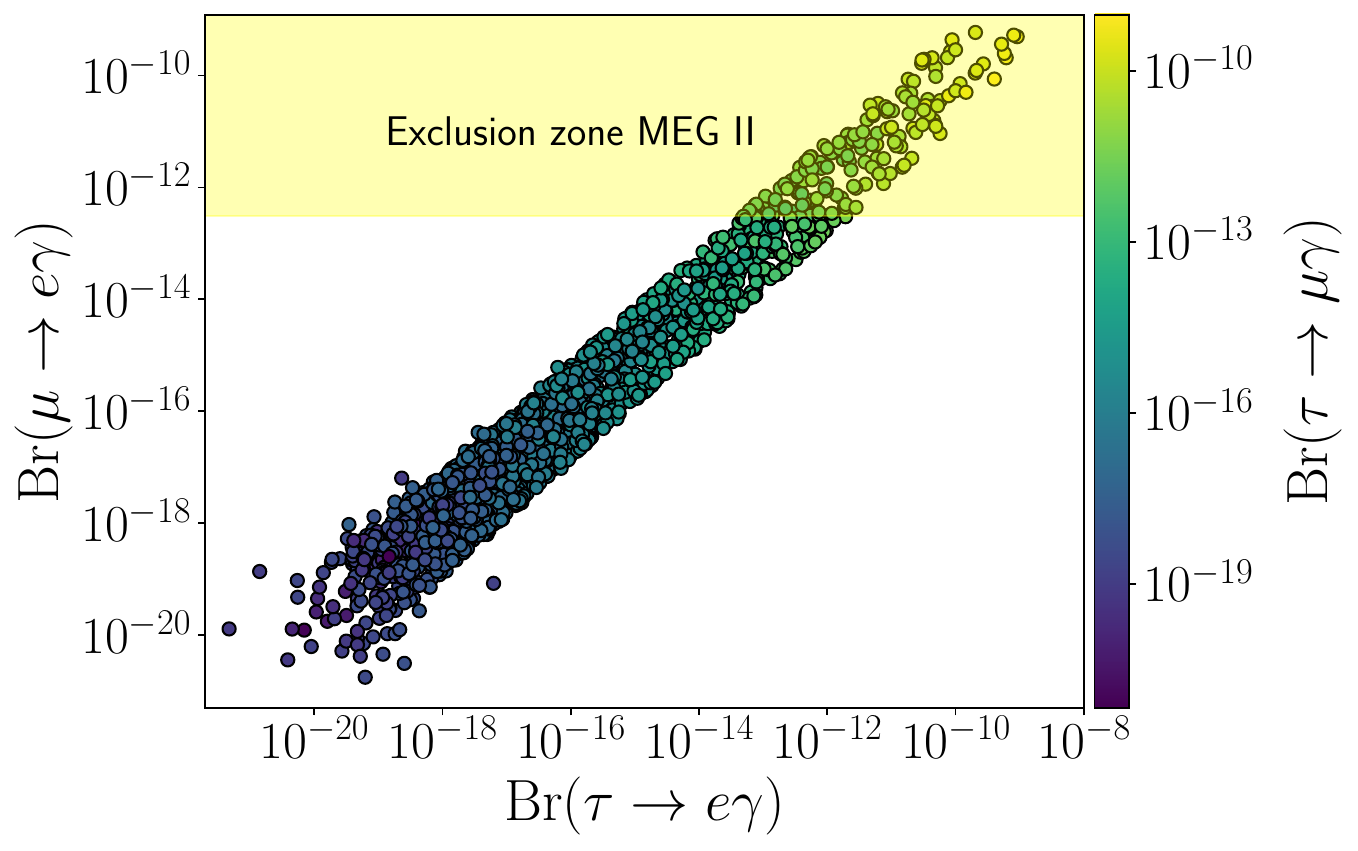}}
\subfigure[]{\includegraphics[width=0.49\linewidth]{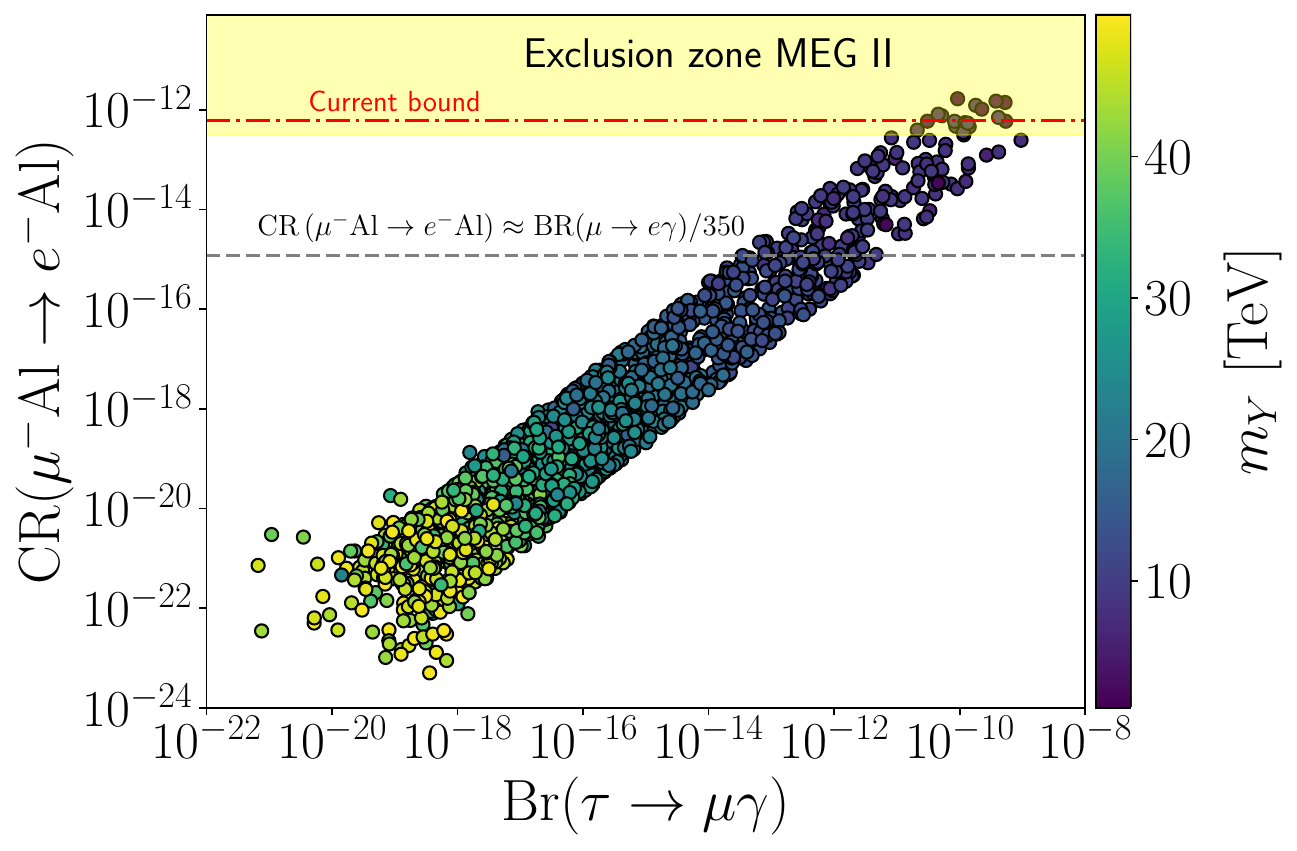}}
\subfigure[]{\includegraphics[width=0.49\linewidth]{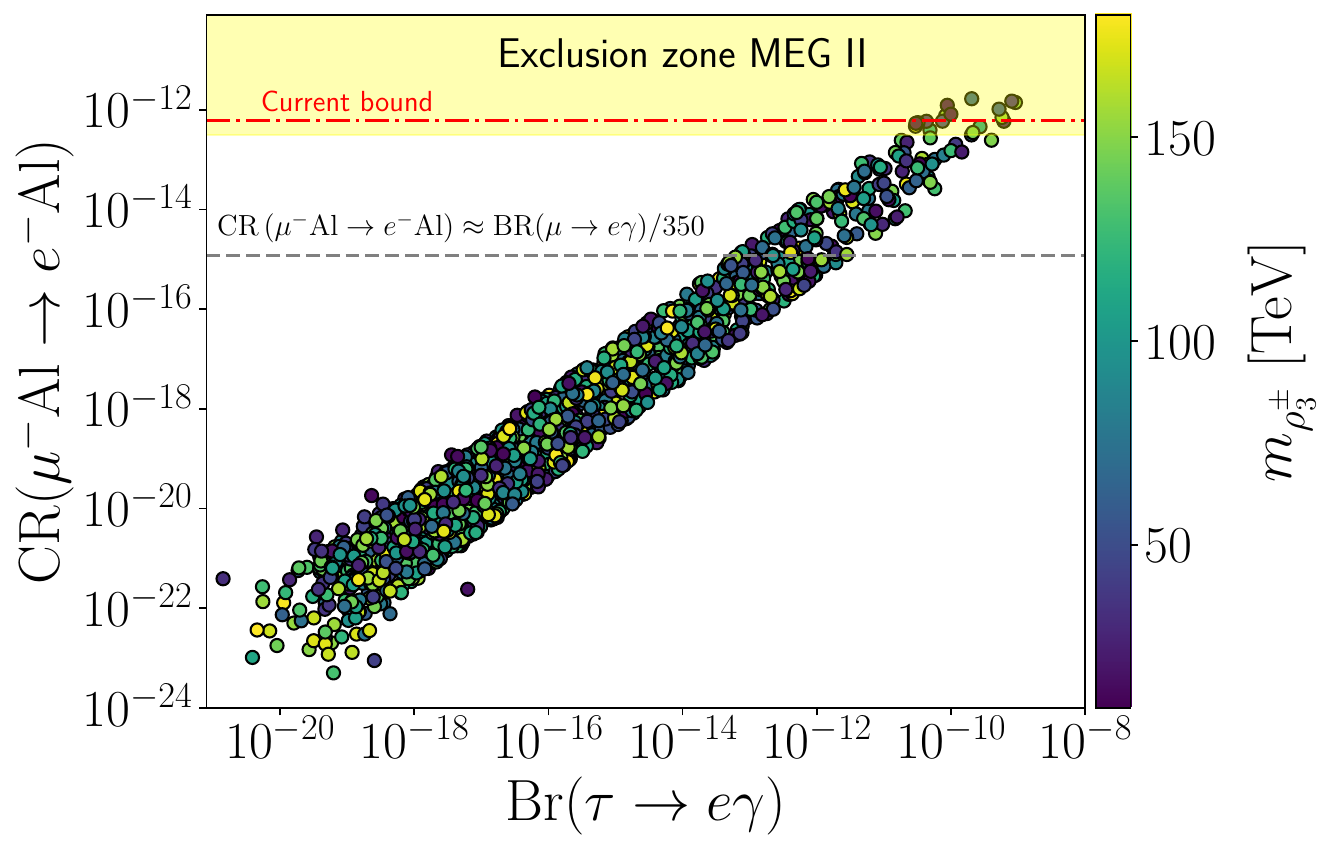}}
\caption{Correlation between the branching ratios $\text{Br}(\mu \to e \gamma)$ and $\text{Br}(\tau \to e \gamma)$ (panel (a)), $\text{Br}(\tau \to \mu \gamma)$ and the conversion rate $\mathrm{CR}(\mu^- \text{Al} \to e^- \text{Al})$ (panel (b)) and $\text{Br}(\tau \to e \gamma)$ and the conversion rate $\mathrm{CR}(\mu^- \text{Al} \to e^- \text{Al})$. The color gradient represents the values of $\text{Br}(\tau \to \mu \gamma)$ in panel (a), the gauge boson mass $m_Y$ in panel (b) and the charged scalar mass $m_{\rho_3^\pm}$ in panel (c). The yellow shaded region corresponds to the exclusion zone from the MEG II experiment \cite{MEGII:2023ltw}, the red dot dashed line indicates the current bound for $\mu \to e$ conversion in titanium (Ti) nuclei \cite{Wintz:1998rp} and the gray dashed line represents the boundary approximated by Eq.~\eqref{eq:aproxBound}.}
\label{fig:clfvall2}
\end{figure}

In Figure~\ref{fig:clfvall2}, the correlations between the branching ratios $\text{Br}(\mu \to e \gamma)$ and $\text{Br}(\tau \to e \gamma)$, as well as between $\text{Br}(\tau \to \mu \gamma)$ and the conversion rate $\mathrm{CR}(\mu \text{Al} \to e \text{Al})$, are shown. In panel (a), the relationship between $\text{Br}(\mu \to e \gamma)$ and $\text{Br}(\tau \to e \gamma)$ is illustrated through a color gradient that indicates the values of $\text{Br}(\tau \to \mu \gamma)$. This panel provides insight into how lepton flavor violation in the muon and tau decays correlates with the respective branching ratios of these processes. 
In panel (b), the correlation between $\text{Br}(\tau \to \mu \gamma)$ and the conversion rate $\mathrm{CR}(\mu^-\text{Al} \to e^- \text{Al})$ is presented, with a color gradient reflecting the values of the gauge boson mass $m_Y$. 
In panel (c), the correlation between $\text{Br}(\tau \to e \gamma)$ and the conversion rate $\mathrm{CR}(\mu^-\text{Al} \to e^- \text{Al})$ is presented, with a color gradient reflecting the values of the charged scaler mass $m_{\rho_3^{\pm}}$. 
This offers a direct indication of the dependence of the conversion rates on the mass of the new gauge boson in our model. The yellow shaded region in both panels marks the exclusion zone from the experimental limits set by the MEG II experiment \cite{MEGII:2023ltw}.

In  Table \ref{tab:mass_br_table}, we present numerical cases for the heavy neutrino masses $m_{n_4}$ to $m_{n_9}$, also 
include the mass of the gauge boson $m_Y$ and the charged scalar mass $m_{\rho_3^\pm}$, both expressed in TeV,
the corresponding branching ratios for processes $\mu \to e \gamma$, $\tau \to e \gamma$ and $\tau \to \mu \gamma$ and the estimated conversion rate $\mathrm{CR}(\mu^- \mathrm{Al} \to e^- \mathrm{Al})$. These values provide insight into the correlations between the LFV processes and the masses of the heavy neutrinos and gauge bosons within the context of our model.

\setlength{\tabcolsep}{4pt}
\begin{table}[htbp]
\centering
\resizebox{\textwidth}{!}{%
\begin{tabular}{cccccccccccc}
\toprule[0.2mm]
\hline
$m_{n_4}$ & $m_{n_5}$ & $m_{n_6}$ & $m_{n_7}$ & $m_{n_8}$ & $m_{n_9}$ & $m_{Y}$ & $m_{\rho_3^{\pm}}$ & $\text{Br}(\mu \to e \gamma)$ & $\text{Br}(\tau \to e \gamma)$ & $\text{Br}(\tau \to \mu \gamma)$ & $\mathrm{CR}(\mu \mathrm{Al} \to e \mathrm{Al})$ \\
\hline \hline
$4.90$ & $3.52$ & $2.81$ & $3.98$ & $3.23$ & $5.57$ & $10.21$ & $118.99$ & $7.14\times 10^{-13}$ & $1.79\times 10^{-13}$ & $1.23\times 10^{-12}$ & $2.04\times 10^{-15}$\\
$5.37$ & $0.95$ & $1.79$ & $3.23$ & $4.63$ & $1.43$ & $12.84$ & $65.78$ & $1.41\times 10^{-13}$ & $1.16\times 10^{-13}$ & $2.16\times 10^{-14}$ & $4.03\times 10^{-16}$\\
$3.35$ & $2.55$ & $3.07$ & $2.95$ & $1.65$ & $5.82$ & $13.28$ & $167.09$ & $1.65\times 10^{-13}$ & $3.57\times 10^{-14}$ & $6.21\times 10^{-14}$ & $4.72\times 10^{-16}$\\
$2.14$ & $4.31$ & $2.74$ & $5.83$ & $4.47$ & $5.48$ & $10.56$ & $173.46$ & $3.16\times 10^{-14}$ & $1.96\times 10^{-13}$ & $3.12\times 10^{-13}$ & $9.04\times 10^{-17}$\\
$1.47$ & $2.34$ & $1.64$ & $4.60$ & $1.78$ & $4.42$ & $11.22$ & $150.23$ & $2.59\times 10^{-14}$ & $4.71\times 10^{-14}$ & $1.61\times 10^{-13}$ & $7.41\times 10^{-17}$\\
$1.01$ & $5.78$ & $4.70$ & $4.29$ & $1.95$ & $5.84$ & $17.61$ & $118.51$ & $9.32\times 10^{-15}$ & $7.66\times 10^{-15}$ & $1.44\times 10^{-15}$ & $2.66\times 10^{-17}$\\
$4.71$ & $3.30$ & $2.58$ & $3.73$ & $1.70$ & $3.47$ & $14.00$ & $177.13$ & $7.98\times 10^{-15}$ & $2.41\times 10^{-14}$ & $2.17\times 10^{-14}$ & $2.28\times 10^{-17}$\\
$2.94$ & $3.44$ & $1.41$ & $4.70$ & $4.79$ & $5.47$ & $15.59$ & $39.34$ & $3.17\times 10^{-16}$ & $4.58\times 10^{-16}$ & $1.95\times 10^{-16}$ & $9.05\times 10^{-19}$\\
$2.40$ & $1.45$ & $5.88$ & $5.92$ & $4.47$ & $2.16$ & $18.42$ & $28.29$ & $6.02\times 10^{-16}$ & $8.22\times 10^{-16}$ & $3.11\times 10^{-15}$ & $1.72\times 10^{-18}$\\
$4.33$ & $4.22$ & $4.41$ & $4.93$ & $2.57$ & $3.17$ & $25.87$ & $63.28$ & $1.86\times 10^{-17}$ & $7.04\times 10^{-17}$ & $7.33\times 10^{-17}$ & $5.30\times 10^{-20}$\\
$1.21$ & $2.22$ & $1.32$ & $4.38$ & $1.35$ & $3.98$ & $23.03$ & $42.50$ & $1.98\times 10^{-17}$ & $5.38\times 10^{-17}$ & $1.46\times 10^{-16}$ & $5.65\times 10^{-20}$\\
$3.19$ & $1.92$ & $1.15$ & $3.89$ & $3.59$ & $1.22$ & $37.17$ & $126.11$ & $5.27\times 10^{-18}$ & $3.43\times 10^{-18}$ & $5.48\times 10^{-19}$ & $1.51\times 10^{-20}$\\
$2.43$ & $2.85$ & $5.25$ & $0.95$ & $3.76$ & $3.65$ & $45.50$ & $34.87$ & $3.27\times 10^{-18}$ & $1.97\times 10^{-18}$ & $5.58\times 10^{-19}$ & $9.36\times 10^{-21}$\\
$4.75$ & $2.02$ & $1.03$ & $4.09$ & $4.58$ & $4.68$ & $45.53$ & $55.09$ & $3.63\times 10^{-19}$ & $6.12\times 10^{-19}$ & $3.25\times 10^{-19}$ & $1.04\times 10^{-21}$\\
$3.99$ & $1.85$ & $3.44$ & $3.75$ & $5.96$ & $4.04$ & $40.51$ & $102.52$ & $3.30\times 10^{-19}$ & $1.87\times 10^{-19}$ & $9.21\times 10^{-19}$ & $9.41\times 10^{-22}$\\
\hline
\bottomrule[0.2mm]
\end{tabular}}
\caption{Branching ratios for the LFV decays. The first to sixth columns present the numerical values of the heavy neutrino masses $m_{n_i}$ in TeV. The seventh and eighth columns show the mass of the gauge boson $m_Y$ and the charged scalar mass $m_{\rho_3^{\pm}}$ in TeV, respectively. The subsequent columns display the branching ratios for the processes $\mu \to e \gamma$, $\tau \to e \gamma$, and $\tau \to \mu \gamma$, respectively. The final column shows the estimated conversion for the process $\mu \mathrm{Al} \to e \mathrm{Al}$.}
\label{tab:mass_br_table}
\end{table}

In what follows we provide a qualitative discussion about the implications of our model in the muon anomalous magnetic moment. It is important to note that the muon anomalous magnetic moment receives several contributions. These include effects from the virtual exchange of heavy neutral and electrically charged gauge bosons paired with their corresponding charged and neutral leptons, as well as contributions from electrically charged scalars and neutrinos. However, these additional contributions are negligible due to the highly suppressed mixing angle between the neutral CP-even component of the $SU(3)_L$ scalar triplet $\rho$ and the other CP-even scalars. As for the contribution from electrically charged scalars and light active neutrinos, our numerical analysis shows that it can account for the observed magnitude of the muon $g−2$ anomaly when the charged scalars are lighter than $400$ GeV. Nevertheless, this contribution is negative and therefore cannot explain the correct sign of the anomaly.  In  order to successfully reproduce the $g-2$ muon anomaly in our model,  its fermion sector has to be enlarged by the inclusion of charged vector like leptons, transforming as singlets under the $SU(3)_C\times SU(3)_L$ gauge symmetry, in a similar way as done in the 331 model with $D_4$ flavor symmetry of Ref. \cite{Hernandez:2021mxo}. Those charged vector like leptons should have the appropiate transformation properties under the discrete groups of our model in such a way to allow them having Yukawa interactions with the SM charged leptons. This will give rise to one loop level contributions to the muon anomalous magnetic moment mediated by charged vector like leptons and neutral scalars running in the internal lines of the loop. Under these conditions the muon $g-2$ anomaly can be successfully accommodated for an appropiate region of parameter space provided that such exotic charged vector like leptons are included in the fermionic spectrum of our model. We do not perform such fermionic extension in this work since recent lattice QCD calculations of the hadronic vacuum polarization
(HVP) contribution of the muon magnetic moment \cite{FermilabLattice:2019ugu,Borsanyi:2020mff,Lehner:2020crt,Aubin:2019usy} lead to a $1.5\sigma$ deviation \cite{Wittig:2023pcl} of its SM prediction from the corresponding experimental value and in addition, a recent study of the hadronic vacuum polarization contribution to the muon $g-2$ at long distances has found that the SM prediction for the total muon anomalous magnetic moment is in agreement with the current experimental average \cite{Djukanovic:2024cmq}.\newline  
Although the model includes sources of CP violation through complex Yukawa couplings and active–sterile neutrino mixing, the contributions to the electron electric dipole moment (EDM) are strongly suppressed. This suppression is due to the large masses of the new charged scalar $\rho_3^\pm$ and the exotic gauge boson $Y^\pm$, which, due to flavour constraints, are typically required to lie above the multi-TeV scale. As pointed out in Ref.~\cite{Crivellin:2018qmi}, the corresponding Wilson coefficients scale inversely with the square of these masses. Furthermore, the structure imposed by the $A_4 \times Z_N$ flavour symmetries tends to align the Yukawa couplings and reduce the number of physical CP-violating phases. Together with the small active–sterile mixing parameters resulting from the bounds imposed by charged lepton flavour-violating processes, it follows that the predicted contributions to the electric dipole moment $d_e$ of the electron are well below the current experimental sensitivity. Therefore, the electron EDM does not place any significant constraint on the model parameter space.\newline
To conclude this section, we present some remarks on the decays of quasi-Dirac sterile neutrinos and compare our predictions for these decays and lepton flavor violating (LFV) signals with those obtained in other models featuring extended gauge symmetries. Similar to the $U^{\prime}(1)$ model discussed in Ref. \cite{Deppisch:2013cya}, our model predicts that sterile neutrinos undergo two-body decays: $N\rightarrow l^{\pm}_iW^{\mp}$, $\nu_iZ$ and $\nu_ih$ (where $i=1,2,3$ is a flavor index). These decay modes are suppressed due to the small active-sterile neutrino mixing angle, $\theta\sim\mathcal{O}(10^{-2})$. The above mentioned two body decays lead to three-body decays: $N\rightarrow l^{+}_il^{-}_j\nu_k$, $N\rightarrow l^{-}_iu_j\bar{d}_k$, $N\rightarrow b\bar{b}\nu_k$ (where $i,j,k=1,2,3$ are flavor indices), which are also present in the $U^{\prime}(1)$ model of Ref. \cite{Deppisch:2013cya}. As a result, we anticipate similar predictions for the total cross section of the LFV process $pp\rightarrow NN\rightarrow e^{\pm}\mu^{\mp}4j$, as well as for sterile neutrino decays, compared to those found in Ref. \cite{Deppisch:2013cya}. However, a slight deviation is expected in the decay rate of  $N\rightarrow l^{+}_il^{-}_j\nu_k$, as our model includes contributions from off-shell $W$ and $W^{\prime}$ gauge bosons  whereas in Ref. \cite{Deppisch:2013cya}, only the off-shell $W$ boson contributes. Notably, the contribution from the off-shell $W^{\prime}$ boson in our model is significantly suppressed by a factor of approximately $\frac{M^4_W}{M^4_{W^{\prime}}}$ relative to the contribution from the $W$ boson. Furthermore, we anticipate similar predictions for sterile neutrino decay rates and the cross section of the LFV process $pp\rightarrow NN\rightarrow e^{\pm}\mu^{\mp}$ to those found in the left-right symmetric model of Refs. \cite{AguilarSaavedra:2012fu,Das:2012ii}. However, the charged lepton flavor violating process $\mu\to e\gamma$ serves as a key discriminator between our model and the left-right symmetric model of Refs. \cite{AguilarSaavedra:2012fu,Das:2012ii}. In our case, the dominant contributions to $\mu\to e\gamma$ originates from  loop diagrams involving $W^{\prime}$ and charged scalar exchanges, whereas in the left-right symmetric models of Refs. \cite{AguilarSaavedra:2012fu,Das:2012ii}, such CLFV process receives contributions from the virtual exchange of the $W_R$ gauge boson and the doubly charged scalars running in the internal lines of the loop.

\section{Conclusions}
\label{conclusions}
We have constructed a $SU(3)_C\times SU(3)_L\times U(1)_X$ model based on the $A_4$ flavor symmetry. This version is different from previous versions by incorporating the $U(1)_{L_g}$ global lepton number symmetry and the $Z_2 \times Z^\prime_2 \times Z_3 \times Z_4 \times Z_7 \times Z_{10}$ discrete group. The model introduces right-handed neutrinos as the third components of the $SU(3)_L$ leptonic triplets and three additional neutral fermions, $N_{aR}$ to generate the light neutrino masses through an inverse seesaw mechanism. The smallness of the $\mu$ parameter of the inverse seesaw is attributed to higher-dimensional Yukawa operators which give rise small lepton number violating Majorana mass terms parameters after the spontaneous breaking of the $U(1)_{L_g}$ global symmetry. \\
The $A_4, Z_3, Z_4$ and $Z_7$ symmetries ensure that the top quark is generated from a renormalizable Yukawa interaction, whereas the masses of lighter fermions arise from non-renormalizable Yukawa interactions after the spontaneous breaking of discrete symmetries. In addition, these discrete symmetries allow for a reduction of fermion sector parameters and yield a hierarchical structure in the entries of the SM charged fermion mass matrices, which effectively explains the SM charged fermion masses and quark mixing pattern. Due to the discrete symmetries of the model, the quark mixing arises only from the up-type quark sector. The $Z_{10}$ symmetry distinguishes the $A_4$ scalar triplets participating in the quark and neutrino Yukawa interactions from the ones appearing in the charged lepton
Yukawa terms, then allowing us to
separate the mixing in the neutrino and quark sectors. The model successfully predicts SM fermion masses and mixing patterns, consistent with experimental observations. The sizable neutrino Yukawa couplings predicted by the model could lead to observable rates of cLFV processes, which could be tested in future experiments.

\section*{Acknowledgments}
The authors are very grateful to L.T. Hue, for his involvement in the initial stages of the project and for very useful discussions. This research has received funding from Fondecyt (Chile), Grants
No.~1210378, No.~1241855, 
No~1210131, ANID PIA/APOYO AFB230003 and Proyecto Milenio-ANID: ICN2019\_044. D.T.Huong acknowledges the financial support of the Vietnam Academy of Science and Technology under Grant No.~NVCC05.05/24-25.

\appendix

\section{The product rules for $A_{4}$} \label{sec:appexA4}

\label{A4}The $A_{4}$ group has one three-dimensional $\mathbf{3}$\ and
three distinct one-dimensional $\mathbf{1}$, $\mathbf{1}^{\prime}$ and $%
\mathbf{1}^{\prime \prime}$ irreducible representations, satisfying the
following product rules:
\bea
&&\hspace{18mm}\mathbf{3}\otimes \mathbf{3}=\mathbf{3}_{s}\oplus \mathbf{3}
_{a}\oplus \mathbf{1}\oplus \mathbf{1}^{\prime}\oplus \mathbf{1}^{\prime
\prime},  \label{A4-singlet-multiplication} \\[0.12in]
&&\mathbf{1}\otimes \mathbf{1}=\mathbf{1},\hspace{5mm}\mathbf{1}^{\prime
}\otimes \mathbf{1}^{\prime \prime}=\mathbf{1},\hspace{5mm}\mathbf{1}
^{\prime}\otimes \mathbf{1}^{\prime}=\mathbf{1}^{\prime \prime},\hspace{ 5mm}%
\mathbf{1}^{\prime \prime}\otimes \mathbf{1}^{\prime \prime}=\mathbf{1}
^{\prime},  \notag
\eea
Considering $\left( x_{1},x_{2},x_{3}\right) $ and $\left(
y_{1},y_{2},y_{3}\right) $ as the basis vectors for two $A_{4}$-triplets $%
\mathbf{3}$, the following relations are fullfilled\textbf{:}

\bea
&&\left( \mathbf{3}\otimes \mathbf{3}\right)_{\mathbf{1} }%
=x_{1}y_{1}+x_{2}y_{2}+x_{3}y_{3},  \label{triplet-vectors} \\
&&\left( \mathbf{3}\otimes \mathbf{3}\right)_{\mathbf{3}_{s}}=\left(
x_{2}y_{3}+x_{3}y_{2},x_{3}y_{1}+x_{1}y_{3},x_{1}y_{2}+x_{2}y_{1}\right) ,\
\ \ \ \left( \mathbf{3}\otimes \mathbf{3}\right)_{\mathbf{1}^{\prime }}%
=x_{1}y_{1}+\om x_{2}y_{2}+\om^{2}x_{3}y_{3},  \crn
&&\left( \mathbf{3}\otimes \mathbf{3}\right)_{\mathbf{3}_{a}}=\left(
x_{2}y_{3}-x_{3}y_{2},x_{3}y_{1}-x_{1}y_{3},x_{1}y_{2}-x_{2}y_{1}\right) ,\
\ \ \left( \mathbf{3}\otimes \mathbf{3}\right)_{\mathbf{1}^{\prime \prime }}%
=x_{1}y_{1}+\om^{2}x_{2}y_{2}+\om x_{3}y_{3},  \notag
\eea
where $\om =e^{i\fr{2\pi}{3}}$. The representation $\mathbf{1}$ is
trivial, while the non-trivial $\mathbf{1}^{\prime}$ and $\mathbf{1}
^{\prime \prime}$ are complex conjugate to each other. Some reviews of
discrete symmetries in particle physics are found in Refs. \cite%
{Ishimori:2010au,Altarelli:2010gt,King:2013eh, King:2014nza}.

\allowdisplaybreaks

\section{\label{app_Higgs} Higgs sector}

The renormalized Higgs potential respecting the whole symmetry group
mentioned in this work is:
\begin{align}  \label{eq_Higgspotential1}
V_H = &\ V_{H,2} +V_{H,3} +V_{H,4},  \crn
V_{H,2}= & \sum_{k=1}^{13} \bar{\mu}_{s_k}^2\left(s_k^{\dagger}s_k%
\right)_{1}+ \left(\mu_{1\zeta \var  }^2 \zeta \var  ^*
+\mu_{\zeta}^2\zeta^2 +\mu_{\zeta \var  }^2\zeta\var   +
\mu_{\eta}^2\eta^2 +\mu_{\Phi \Xi}^2\Xi^* \Phi +\mu_{\var  }^2
\var  ^2 +\mathrm{H.c.}\right)_{1},  \crn
V_{H,3}= & \ \eta \left(\eta^2 \la _{\eta} + \la _{\eta}^{(1)} \eta\eta^*
+ \la _{\eta\zeta}\zeta^2 +\la _{\eta \zeta}^{(1)} \zeta\zeta^*
+\la _{\eta \zeta}^{(2)} \zeta^{*2} + \la _{\eta\zeta\var  }
\zeta\var   + \la _{\eta \zeta \var  }^{(1)} \zeta\var  ^*
+\la _{\eta\zeta \var  }^{(2)} \zeta^* \var   \right.  \crn
&\left. +\la _{\eta \zeta \var  }^{(3)} \zeta^* \var  ^* +
\la _{\eta \xi} \xi\xi^* + \la _{\eta\Xi} \Xi\Xi^* +\la _{\eta
\Phi} \Phi^* \Phi +\la _{\eta \Phi \Xi} \Xi^* \Phi +
\la _{\eta\Phi\Xi}^{(1)} \Xi\Phi^* \right.  \crn
&\left. +\la _{\eta \var  } \var  ^2 +\la _{\eta
\var  }^{(1)} \var  ^* \var   +\la _{\eta \phi} \phi^*\phi
\right) +\la _{\eta \xi \va }\left(\eta\xi\right)_{1^{\prime }}
\va ^* +\la _{\eta \xi \va }^{(1)}
\left(\eta\xi^*\right)_{1^{\prime \prime }} \va  +\mathrm{H.c.},  \notag
\\
V_{H,4}= &  \sum_{k=1}^{13}\bar{\la }_{s_k}\left(s_k^{\dagger}s_k\right)^2
+ \sum_{k,l=1,l<k}^{13}\bar{\la }_{s_ks_l}\left(
s_k^{\dagger}s_k\right)\left( s_l^{\dagger}s_l\right) + \bar{\la }%
_{2\chi\rho}\left(\chi^{\dagger}\rho\right)\left(\rho^{\dagger}\chi\right)
\crn
&+ \sum_{k=1,2} \left[ \left(s_k^{\dagger}s_k\right) \left( \la _{s_k
\zeta} \zeta^2 + \la _{s_k \zeta\var  }\zeta \var   + \la _{s_k
\zeta\var  }^{(1)} \zeta\var  ^* + \la _{s_k\eta}\eta^2
+\la _{s_k \Phi \Xi} \Xi^* \Phi +\la _{s_k \var  }
\var  ^2\right)_{1} +\mathrm{H.c.}\right],  \crn
& +V_{H,4}^{(1)} +V_{H,4}^{(2)},  \crn
V_{H, 4}^{(1)} = & \ \zeta^4 \la _{\zeta} +\zeta^3
\left(\la _{\zeta}^{(1)} \zeta^* +\la _{\zeta\var  } \var
+\la _{\zeta \var  }^{(1)} \var  ^*\right) +\zeta
\left(\la _{\zeta \var  }^{(10)} \var  ^{*2} \var
+\la _{\zeta \var  }^{(11)} \var  ^{*3} +\la _{\zeta
\var  }^{(7)} \var  ^3 +\la _{\zeta \var  }^{(8)} \zeta^*
\var  ^2 +\la _{\zeta \var  }^{(9)} \var  ^* \var  ^2\right)
\crn
&+\zeta^2 \left[ \la _{\zeta \var  }^{(2)} \var  ^2 +\zeta^*
\left(\la _{\zeta \var  }^{(3)} \var   +\la _{\zeta
\var  }^{(5)} \var  ^*\right) +\la _{\zeta \var  }^{(4)}
\var  ^* \var   +\la _{\zeta \var  }^{(6)} \var  ^{*2}\right]
+\la _{\var  } \var  ^4 +\la _{\var  }^{(1)} \var  ^*
\var  ^3  \crn
&+ \la _{\zeta \var   \Phi \Xi} \zeta\Xi^* \Phi \var   +
\la _{\zeta \var   \Phi \Xi}^{(1)} \zeta \Xi\Phi^* \var   +
\la _{\zeta \var   \Phi \Xi}^{(2)} \zeta\Xi^* \var  ^* \Phi +
\la _{\zeta \var   \Phi \Xi}^{(3)} \zeta \Xi\Phi^* \var  ^*  \notag
\\
&+\phi \left( \la _{\zeta \phi \Phi \Xi} \zeta \Xi\Phi +
\la _{\var  \phi\Phi \Xi} \Xi\Phi \var   +\la _{\phi \Xi}^{(1)}
\Xi^{*3} +\la _{\phi \Phi}^{(1)} \Phi^{*3}\right)  \crn
&+\phi^* \left( \la _{\zeta \phi \Phi \Xi}^{(1)} \zeta \Xi^* \Phi^*
+\la _{\var  \phi \Phi\Xi}^{(1)} \Xi^* \Phi^* \var   \right) +
\la _{\eta}\eta^4 +\la _{\eta}^{(1)} \eta^3\eta^* +\phi^3
\left(\la _{\phi\Xi}\Xi +\la _{\phi \Phi} \Phi \right)  \crn
& +\xi \xi^* \left( \la _{\zeta \xi}\zeta^2 + \la _{\zeta \var
\xi} \zeta\var   + \la _{\zeta \var   \xi}^{(1)} \zeta\var  ^*
+\la _{\Phi \Xi \xi} \Xi^* \Phi +\la _{\var  \xi} \var  ^2\right)
\crn
& +\eta^2 \left( \la _{\eta \zeta}\zeta^2 + \la _{\eta \zeta}^{(1)}
\zeta\zeta^* +\la _{\eta\zeta}^{(2)} \zeta^{*2} + \la _{\eta \zeta
\var  } \zeta\var   + \la _{\eta\zeta\var  }^{(1)}
\zeta\var  ^* +\la _{\eta \zeta \var  }^{(2)} \zeta^* \var
+\la _{\eta\zeta \var  }^{(3)} \zeta^* \var  ^* + \la _{\eta\xi}
\xi^*\xi \right.  \crn
&\left. \quad \quad + \la _{\eta \Xi} \Xi\Xi^* +\la _{\eta \Phi}
\Phi^* \Phi +\la _{\eta\Phi \Xi} \Xi^* \Phi + \la _{\eta
\Phi\Xi}^{(1)} \Xi\Phi^* +\la _{\eta \var  } \var  ^2
+\la _{\eta \var  }^{(1)} \var  ^* \var   +\la _{\eta
\var  }^{(2)} \var  ^{*2}+\la _{\eta \phi} \phi^* \phi \right)
\crn
&+\eta \eta^* \left( \la _{\eta \zeta}^{(3)}\zeta^2 +
\la _{\eta\zeta\var  }^{(4)} \zeta\var   + \la _{\eta\zeta
\var  }^{(5)} \zeta\var  ^* +\la _{\eta \Phi \Xi}^{(2)} \Xi^* \Phi
+\la _{\eta \var  }^{(3)} \var  ^2\right)  \crn
&+\zeta^2 \left( \la _{\zeta \Xi} \Xi \Xi^* +\la _{\zeta \Phi} \Phi^*
\Phi +\la _{\zeta\Phi \Xi} \Xi^* \Phi +\la _{\zeta \Phi \Xi}^{(1)} \Xi
\Phi^*\right)  \crn
& +\zeta \left[\la _{\zeta \Phi \Xi}^{(2)} \zeta^* \Xi^* \Phi +\var
\left( \la _{\zeta \var  \Xi} \Xi \Xi^* +\la _{\zeta \var
\Phi} \Phi^* \Phi \right) +\var  ^* \left(\la _{\zeta \var
\Xi}^{(1)} \Xi\Xi^* +\la _{\zeta\var   \Phi}^{(1)} \Phi^* \Phi\right) %
\right]  \crn
& +\phi \phi^* \left[ \left(\la _{\zeta \var   \phi} \var
+\la _{\zeta \var  \phi}^{(1)} \var  ^*\right)\zeta +
\la _{\zeta \phi}\zeta^2 +\la _{\var   \phi} \var  ^2
+\la _{\phi \Phi \Xi}\Xi^* \Phi \right]  \crn
&+\phi^2 \left[\zeta \left(\la _{\zeta \phi \Xi} \Xi^* +\la _{\zeta
\phi \Phi} \Phi^*\right) +\var   (\la _{\var   \phi \Xi} \Xi^*
+\la _{\var   \phi \Phi} \Phi^*)\right] +\phi^{*2} \left[\zeta \left(
\la _{\zeta \phi \Xi}^{(2)}\Xi +\la _{\zeta \phi \Phi}^{(2)} \Phi
\right) +\var   \left( \la _{\var   \phi\Xi}^{(2)}\Xi
+\la _{\var   \phi \Phi}^{(2)} \Phi \right)\right]  \crn
& +\phi \left[\zeta \left( \la _{\zeta \phi \Xi}^{(1)}\Xi^2
+\la _{\zeta \phi \Phi}^{(1)} \Phi^2\right) +\var   \left(
\la _{\var   \phi \Xi}^{(1)}\Xi^2 +\la _{\var   \phi \Phi}^{(1)}
\Phi^2\right) +\la _{\phi \Phi \Xi}^{(1)} \Xi^* \Phi^{*2} +\la _{\phi
\Phi \Xi}^{(2)} \Xi^{*2} \Phi^*\right]  \crn
& +\phi^* \left[\zeta \left(\la _{\zeta \phi \Xi}^{(3)} \Xi^{*2}
+\la _{\zeta \phi \Phi}^{(3)} \Phi^{*2}\right) +\var
\left(\la _{\var   \phi\Xi}^{(3)} \Xi^{*2} +\la _{\var  \phi
\Phi}^{(3)} \Phi^{*2}\right)\right] +\la _{\Phi \Xi} \Xi^* \Phi^* \Phi^2
+\la _{\Phi \Xi}^{(1)} \Xi^{*2} \Phi^2 + \la _{\Phi\Xi}^{(2)}
\Xi\Xi^{*2} \Phi  \crn
& +\var  ^2 \left( \la _{\var   \Xi} \Xi\Xi^*
+\la _{\var  \Phi} \Phi^* \Phi +\la _{\var   \Phi \Xi} \Xi^*
\Phi + \la _{\var  \Phi\Xi}^{(1)} \Xi\Phi^*\right)
+\la _{\var  \Phi \Xi}^{(2)} \Xi^* \Phi \var  ^*\var   +\mathrm{%
H.c.},  \crn
V_{H, 4}^{(2)} = & \sum_{k=3}^5s_k s_k^* \left( \la _{s_k\zeta}\zeta^2 +
\la _{s_k\zeta \var  } \zeta\var   + \la _{s_k\zeta
\var  }^{(1)} \zeta\var  ^* + \la _{s_k\eta}\eta^2
+\la _{s_k\Phi \Xi} \Xi^* \Phi +\la _{s_k\var  }
\var  ^2\right)_{1}  \crn
&+\va \va ^* \left( \la _{\zeta \va }\zeta^2 + \la _{\zeta
\var   \va } \zeta\var   + \la _{\zeta \var   \va }^{(1)}
\zeta\var  ^* + \la _{\eta \va }\eta^2 +\la _{\Phi\Xi\va }
\Xi^* \Phi +\la _{\var   \va } \var  ^2\right)_{1} +
\la _{\xi\va }^{(1)} \left(\xi^2\right)_{1^{\prime \prime
}}\va ^{*2}  \crn
&+ \va ^* \xi \left( \la _{\zeta \xi \va }\zeta^2 +\la _{\zeta
\xi \va }^{(2)} \zeta\zeta^* +\la _{\zeta\var   \xi\va }
\zeta\var   +\la _{\zeta \var   \xi \va }^{(2)} \zeta\var  ^*
+ \la _{\eta \xi \va }\eta^2 + \la _{\eta \xi \va }^{(2)}
\eta\eta^* + \la _{\Xi \xi \va } \Xi\Xi^* +\la _{\Phi \Xi \xi
\va }\Xi^* \Phi \right.  \crn
&\left. +\la _{\Phi \xi \va } \Phi^* \Phi +\la _{\var   \xi
\va } \var  ^2 +\la _{\var   \xi \va }^{(2)} \var  ^*
\var   +\la _{\phi \xi \va } \phi^* \phi \right) + \la _{\xi
\va } \va ^* \xi^2\xi^*  \crn
&+\va  \xi^* \left(\la _{\zeta \xi \va }^{(1)}\zeta^2 +
\la _{\zeta\var   \xi \va }^{(1)} \zeta\var   +
\la _{\zeta\var   \xi \va }^{(3)} \zeta\var  ^* + \la _{\eta
\xi\va }^{(1)}\eta^2 +\la _{\Phi \Xi \xi \va }^{(1)} \Xi^* \Phi
+\la  _{\var   \xi \va }^{(1)} \var  ^2\right) +\mathrm{H.c.},
\end{align}
where $s_k$ runs over $A_4$ Higgs representations introduced in this model, $%
s_k=\chi ,\rho ,\si _1,\si _2,\si _3,\eta ,\zeta ,\var   ,\phi
,\Phi ,\Xi ,\xi ,\va $ corresponding to the orders appearing in Table~\ref%
{tab:scalars}: $k=1,2,3,...,13$. The $SU(3)_L$ Higgs triplets $\chi$ and $%
\rho$ only appear in $V_{H,2}$ and the two first lines of $V_{H,4}$. In
contrast, $V_{H,3}$, $V_{H,4}^{(1)}$, and $V_{H,4}^{(2)}$ consist of only $%
SU(3)_L$ singlets. Except two terms with $A_4$ singlets $\va $ and $%
\va ^*$, all terms in $V_{H,3}$ are products of three $A_4$ triplets. The
Lagrangian part $V^{(1)}_{H,4}$ consists of all products of four $A_4$
triplets. The first and second lines of $V^{(2)}_{H,4}$ consist of products
of two $A_4$ singlets with two $A_4$ triplets. The remaining lines consist
of products of one $A_4$ singlet with three $A_4$ triplets

The expansion rule for determining invariant terms of a product of three $%
A_4 $ triplets $s_is_js_k$ is
\be  \label{eq_3A4triplets}
\la _x s_is_js_k \equiv \la _{x,1} \left[\left(s_is_j\right)_{3_s}
s_k\right]_{1} +\la _{x,2} \left[\left(s_is_j\right)_{3_a} s_k\right]_{1},
\ee
where $\la _x$, $\la _{x,1}$, and $\la _{x,2}$ are couplings,
implying that every term having three $A_4$ triplets in $V_{H,3}$ and $%
V^{(2)}_{H,4}$ stands for two independent terms, in general. Note that $%
\left(s_is_i\right)_{3_a}=0$, hence the first term automatically vanish if $%
i=j$.

The expansions of products of the four reps. 3 of the $A_4$ symmetry for
every term $\la  s_is_js_ks_l$ in $V^{(1)}_{H,4}$ are defined generally
as:
\begin{align}  \label{eq_A4sijkl}
\left[ \la  (s_is_j)(s_ks_l)\right]_1&\equiv \la _1
(s_is_j)_1(s_ks_l)_1 + \la _2 (s_is_j)_{1^{\prime }}(s_ks_l)_{1^{\prime
\prime }} + \la _3 (s_is_j)_{1^{\prime \prime }}(s_ks_l)_{1^{\prime }}
+\la _4 \left[ (s_is_j)_{3s}(s_ks_l)_{3s}\right]_1  \crn
&+\la _5 \left[ (s_is_j)_{3s}(s_ks_l)_{3a}\right]_1 +\la _6 \left[
(s_is_j)_{3a}(s_ks_l)_{3s}\right]_1 +\la _7 \left[
(s_is_j)_{3a}(s_ks_l)_{3a}\right]_1,
\end{align}
where $\la $ is the coupling appearing in the Higgs potential~%
\eqref{eq_Higgspotential1}, while $\la _{i}$ with $i=1,2,3,...,7$ are the
independent couplings corresponding to the seven $A_4$ invariant terms. For
the Higgs sector introduced in this work, the products like $\left[
(s_is_k)(s_is_l)\right]_1$, $\left[ (s_is_l)(s_js_k)\right]_1$,... are not
included in the Higgs potential because we can prove that they are always
written in terms of linear combinations of the seven $A_4$ products given in
Eq.~\ref{eq_A4sijkl}.

Let us note that due to the antisymmetry and symmetry properties of the $\textbf{3}_a$
and $\textbf{3}_s$ triplet components in the products $(s_is_i)$, we obtain $%
(\textbf{3}_s\textbf{3}_a)_1 +\mathrm{H.c.}=0$. Hence many terms having this invariance does
not appear in the Higgs potential.

With the Higgs potential given above, the VEV patterns satisfy the all the
equations corresponding to the minimal conditions of the Higgs potential.

The masses and mixing parameters of the singly charged Higgs boson are
computed in the general cases of the Higgs potential, the minimal equation
automatically satisfies $\fr{\partial V_H}{\partial \chi^0_1}=0$ at $%
\chi^0_1=\langle\chi^0_1\rangle=0$. Using two other equations $\fr{%
\partial V_H}{\partial\chi_3}=0$ and $\fr{\partial V_H}{\partial\rho_2}=0$
at the corresponding VEVs, we write $\bar{\mu}^2_{\chi}$ and $\bar{\mu}%
^2_{\rho}$ as the functions of other parameters in the Higgs potential as
follows:
\begin{align}  \label{eq_minimal23}
\bar{\mu}_{\chi}^2=& c_{\ga }^2\left(-2 c_{\al }^2 v_{\zeta}^2\la {%
\chi\zeta} -\bar{\la }_{\chi\zeta} v_{\zeta}^2 -\bar{\la }_{\chi\Xi}
v_{\Xi}^2\right) -c_{\ga } s_{\al } s_{\ga } v_{\zeta}
v_{\var  }\la {\chi\zeta\var  } -\fr{\bar{\la }_{1\chi\rho}
v_{\rho}^2}{2} -\bar{\la }_{\chi} v_{\chi}^2 +v_{\eta}^2 (-\bar{\la }%
_{\chi\eta}-2\la {\chi\eta}) -\bar{\la }_{\chi\xi} v_{\xi}^2  \crn
& -\fr{\bar{\la }_{\chi\sigma1} v_{\sigma1}^2}{2} -\fr{\bar{\la }%
_{\chi\sigma2} v_{\sigma2}^2}{2} -\fr{\bar{\la }_{\chi\sigma3}
v_{\sigma3}^2}{2} -\fr{\bar{\la }_{\chi\va } v_{\va }^2}{3}
+s_{\ga }^2\left(-\bar{\la }_{\chi\Phi} v_{\Phi}^2-\bar{\la }%
_{\chi\var  } v_{\var  }^2\right) -\bar{\la }_{\chi \phi}
v_{\phi}^2,  \crn
\bar{\mu}_{\rho}^2=& c_{\ga }^2\left(-2 c_{\al }^2 v_{\zeta}^2\la {%
\rho\zeta} -\bar{\la }_{\rho\zeta} v_{\zeta}^2 -\bar{\la }_{\rho\Xi}
v_{\Xi}^2\right) -c_{\ga } s_{\al } s_{\ga } v_{\zeta}
v_{\var  }\la {\rho\zeta\var  } -\fr{\bar{\la }_{1\chi\rho}
v_{\chi}^2}{2} -\bar{\la }_{\rho} v_{\rho}^2 +v_{\eta}^2(-\bar{\la }%
_{\rho\eta} -2\la {\rho\eta}) -\bar{\la }_{\rho\xi} v_{\xi}^2  \notag
\\
&-\fr{\bar{\la }_{\rho\sigma1} v_{\sigma1}^2}{2} -\fr{\bar{\la }%
_{\rho\sigma3}v_{\sigma2}^2}{2} -\fr{\bar{\la }_{\rho\sigma3}
v_{\sigma3}^2}{2} -\fr{\bar{\la }_{\rho\va } v_{\va }^2}{3}
+s_{\ga }^2\left(-\bar{\la }_{\rho\Phi} v_{\Phi}^2 -\bar{\la }%
_{\rho\var  } v_{\var  }^2\right) -\bar{\la }_{\rho \phi}
v_{\phi}^2.
\end{align}
Inserting these into the Higgs potential, we easily find that $\rho^{\pm}_1$
are exactly the Goldstone bosons absorbed by the SM charged gauge bosons $%
W^{\pm}$. On the other hand the two components $\chi^-_2$ and $\rho^-_3$ mix
with each other to result in a Goldstone boson $G^-_Y$ of the heavy charged
gauge boson $Y^-$ and a physical charged Higgs boson $h^-$. The mixing and
masses are
\begin{align}  \label{eq_Cpm}
\begin{pmatrix}
\chi^{\pm}_2 \\
\rho^{\pm}_3%
\end{pmatrix}
= C_{\pm}
\begin{pmatrix}
G^{\pm}_Y \\
h^{\pm}%
\end{pmatrix}%
, \quad C_{\pm}=\left(
\begin{array}{cc}
c_x & s_x \\
-s_x & c_x \\
\end{array}
\right), \quad m^2_{h^{\pm}}=\fr{\bar{\la  }_{2\chi \rho }}{2}
\left(v_{\rho }^2+v_{\chi }^2\right),
\end{align}
where $c_x\equiv \cos\theta_x$, $s_x\equiv\sin\theta_x$, and
\be  \label{eq_tx}
t_x\equiv \tan\theta_x=\fr{v_{\rho}}{v_{\chi}}.
\ee

\bibliographystyle{utphys}
\bibliography{Biblio331A4June2022}

\end{document}